\documentclass[a4paper,12pt]{article}
\usepackage{graphics}
\usepackage{epsf,epsfig,psfrag,color,epstopdf}
\usepackage{mathrsfs}
\usepackage{latexsym}
\usepackage{ulem}
\usepackage{subfigure}
\usepackage{amsfonts,amssymb,amscd,amsxtra,amsthm,amsmath}
\usepackage{bm}
\usepackage{amsfonts}
\usepackage{dsfont}
\usepackage{multirow}
\usepackage{appendix}
\usepackage{slashed}
\usepackage[active]{srcltx}

\makeatletter
\def\@xfootnote[#1]{%
  \protected@xdef\@thefnmark{#1}%
  \@footnotemark\@footnotetext}
\makeatother

\begin{document}
\def\beq{\begin{equation}}
\def\eeq{\end{equation}}
\def\bea{\begin{eqnarray}}
\def\eea{\end{eqnarray}}
\def\ve{\vert}
\def\vel{\left|}
\def\ver{\right|}
\def\ar{&+& \!\!\!}
\def\nnb{\nonumber}
\def\ga{\left(}
\def\dr{\right)}
\def\aga{\left\{}
\def\adr{\right\}}
\def\rar{\rightarrow}
\def\nnb{\nonumber}
\def\la{\langle}
\def\ra{\rangle}
\def\ba{\begin{array}}
\def\ea{\end{array}}
\def\tep{$B \rar K \ell^+ \ell^-$}
\def\tepm{$B \rar K \mu^+ \mu^-$}
\def\tept{$B \rar K \tau^+ \tau^-$}
\def\ds{\displaystyle}
\title{{\small {\bf Rare $\Lambda_{b}\to \Lambda\ell^{+}\ell^{-}$ decay in the two-Higgs doublet model of
type-III}}}
\author{\vspace{0.01cm}\\
{\small R. F. Alnahdi,~T. Barakat \thanks {electronic address:
tbarakat@ksu.edu.sa},~and H. A. Alhendi}\\
{\small Physics Department, King Saud University}\\
{\small P. O. Box 2455, Riyadh 11451, Saudi Arabia}}
\begin{titlepage}
\maketitle
\thispagestyle{empty}
\begin{abstract}
\baselineskip 0.5 cm
The rare exclusive dileptonic $\Lambda_{b}\to \Lambda\ell^{+}\ell^{-}$ $(\ell =\mu, \tau)$ decays are investigated in the general two-Higgs-doublet model of type III.  A significant enhancement to the branching ratios, differential branching ratios, leptons forward-backward
asymmetry, and the $\Lambda$ baryon polarizations over the standard model is obtained. Measurements of these quantities will be useful for establishing the two-Higgs doublet model.
\\
\vspace{0.5cm} PACS numbers: 12.15.Ji; 12.60.-i\\
\end{abstract}
\end{titlepage}

\section{{\small Introduction}}
\baselineskip .8cm \hspace{0.6cm}
In 2012, the ATLAS and CMS experiments at CERN (Run 1) reported evidence of a particle consistent with the Higgs boson at a mass of
$\sim 125$ GeV [1-5]. This result represents a truly fundamental discovery which is in the right direction at least to understand better the electroweak symmetry breaking via the Higgs mechanism implemented in the standard model (SM) through one scalar $SU(2)_{L}$ doublet. With this discovery the large Hadron collider (LHC) completed the particle content of the SM. Nonetheless, an obvious question we are now facing is whether the discovered $\sim 125$ GeV state corresponds to the SM Higgs boson, or it is just the first signal of a much richer scenario of electroweak symmetry breaking mechanism.

This result initiated physicists to ruminate the different possibilities to search for new physics beyond the SM. One of the most promising scenarios for new physics beyond the SM, is an extended Higgs sector which has rich phenomenology [6]. In this regard, the flavour-changing neutral currents (FCNC) processes, such as the electro-weak penguin decays $b\rightarrow  s\ell ^{+}\ell ^{-}$ is one of these phenomenons [7], and references therein.

Currently, the main interest is focused on the semi-leptonic decays of heavy hadrons $\Lambda_{b}\to \Lambda\ell^{+}\ell^{-}$ which offer cleaner probes compared to non-leptonic exclusive hadronic decays, and give valuable insight into the nature of FCNC. These decays are forbidden at the tree level in
the SM, and they only appear at the one-loop level. Therefore, the study of these rare decays provide sensitive tests of many new physics models beyond the SM. The new physics effects in these decays can appear either by introducing new intermediate particles and interactions into the Wilson coefficients, or through introducing new operators into the effective Hamiltonian of such decays.

As a matter of fact, the exclusive rare $B$ mesons described by FCNC $b\rightarrow  s\ell ^{+}\ell ^{-}$ decays at the quark level, have been extensively studied with varying degrees of theoretical rigor and emphasis. In spite of the progress in the exclusive rare $B$ mesons, so far, we have not seen yet any clear sign of new physics
in this $b$ sector, but there was a tension with the SM predictions in some $b\to s$ penguin induced transitions. For example, the measurement by LHCb collaboration shows several significant deviations on angular observables related to $B\to K^{*}\mu^{+}\mu^{-}$ channels from their corresponding SM expectations [8-18].

Therefore, now, it is of the utmost importance to study any other such semileptonic decay modes in another sector to clarify this situation, and point out the source of these deviations. In this context, the FCNC processes in baryonic sector
receive special attention to search for new physics effects besides the direct searches at LHC.

Apparently, the investigations of FCNC transitions for the bottom baryonic decays can represent a useful ground to find the helicity structure of the effective Hamiltonian which is lost in the hadronization in the meson case [19].

In the last few years, several theoretical works have emerged to better understand $\Lambda_{b}\to \Lambda\ell^{+}\ell^{-}$ $(l =e, \mu, \tau)$ decays in both the SM and beyond [20, 21], and the references therein. On the experimental side, the
first experimental result on rare baryonic decay mode $\Lambda_{b}\to \Lambda\mu^{+}\mu^{-}$ has recently been
reported by the CDF collaboration at Fermilab BR($\Lambda_{b}\to \Lambda\mu^{+}\mu^{-}$)= $[1.73\pm 0.42(stat)\pm 0.55(syst)]\times 10^{-6}$ [22], and LHCb collaboration at CERN has also reported on this branching ratio mode BR($\Lambda_{b}\to \Lambda\mu^{+}\mu^{-}$)= $[0.96\pm 0.16(stat)\pm 0.13(syst)\pm 0.21(norm)]\times 10^{-6}$ [23].

Quite recently, the LHCb Collaboration has reported on both the differential branching ratio of the rare decay $\Lambda_{b}\to \Lambda\mu^{+}\mu^{-}$ and the lepton forward-backward asymmetry ($A_{FB}$) in the dilepton invariant mass-squared region $15 < q^2 < 20~GeV^{2}$ as $dBR(\Lambda_{b}\to \Lambda\mu^{+}\mu^{-})/dq^2$ =$ \left(1.18^{+0.09}_{-0.08}\pm 0.03 \pm 0.27\right)\times 10^{-7}~GeV^{-2}$, and $A_{FB}=-0.05\pm 0.09(stat)\pm 0.03(syst)$ [24]. The errors are still quite large, but one hopes to have more new results in the near future.

Consequently, a deeper understanding of such rare baryonic decays is now entering a new era. One of the motivated scenarios for new physics beyond the SM, is two-Higgs doublet model (2HDM). Basically, the 2HDM has two complex Higgs doublets, $\Phi_{1}$ and $\Phi_{2}$ rather than one, as in the SM, and the 2HDM allows FCNC at tree level, which can be avoided by imposing
an ad hoc discrete symmetry. One of the possibilities to avoid the FCNC is to couple all the quarks
to $\Phi_{2}$, whereas, $\Phi_{1}$ does not couple to quarks at all, which is often known as type I. The second possibility
is to couple $\Phi_{1}$ to the down-type quarks, while $\Phi_{2}$ to couple the up-type quarks, which is known as type II [25].

At the same time, there have been further works on a more general 2HDM without discrete symmetries as in types I and II called  type III. In this type both $\Phi_{1}$ and $\Phi_{2}$ couple to all quarks, and FCNC exists in type III at tree level [26]. It implies that, type III should be parameterized in such a way to suppress the tree-level FCNC couplings of the first two generations
while the tree-level FCNC couplings involving the third generation can be made nonzero as long as they do not violate the existing experimental data, like, $B^{0}-\overline{B^{0}}$ mixing.

In this work, we shall investigate the rare exclusive $\Lambda_{b}\to \Lambda\ell^{+}\ell^{-}$ decays within 2HDM of type III. With this in mind, the structure of this work is organized as follows. In Section 2, we present the effective Hamiltonian for $ b\rightarrow s\ell^{+}\ell^{-}$ transition in the 2HDM. Section 3, contains the parametrization of the matrix elements, and the derivation of the amplitude of $\Lambda_{b}\to \Lambda\ell^{+}\ell^{-}$ decay, as well as, other physical observables like decay rate, leptons forward-backward
asymmetry ($A^{\ell}_{FB}$), and polarization asymmetries of $\Lambda$ baryon in 2HDM of type III. Section 4, is
devoted to the numerical analysis of these observables. Finally, section 5, contains our brief summary and concluding remarks.

\section{{\small The effective Hamiltonian for $b\rightarrow s\ell^{+}\ell^{-}$ transition}}
\hspace{0.6cm} The baryonic $\Lambda_{b}\to \Lambda\ell^{+}\ell^{-}$ $(\ell =\mu, \tau)$ decays at quark level are described by FCNC $b\rightarrow  s\ell ^{+}\ell ^{-}$ transition. The effective Hamiltonian representing these decays in both SM and 2HDM can be written in terms of a set of local operators, and takes the following basic form [27]:
\begin{equation}
H_{eff}=-\frac{4G_{F}}{\sqrt{2}}V_{tb}V^{* }_{ts} \left \{  \sum\limits_{i=1}^{10}C_{i}(\mu)O_{i}(\mu)+\sum\limits_{i=1}^{10}C_{Qi}(\mu)Q_{i}(\mu)\right\},
\end{equation}
where, $G_{F}$ is the Fermi coupling constant, $V_{tb}^{\ast }V_{ts}$ is the relevant Cabibbo-Kobayashi-Maskawa (CKM) matrix elements, and of course the terms proportional to $V_{ub}V_{us}^{*}$ are ignored
since  $|V_{ub}V_{us}^{*}/V_{tb}V_{ts}^{*}|<0.02$. $O_{i}({\mu })$ are the set of the relevant local operators, and $C_{i}({\mu })$ are the Wilson coefficients that describe the short and long distance contributions renormalized at
the energy scale $\mu$ which is usually taken to be the b-quark mass for b-quark decays. The additional operators $Q_{i}({\mu })$ are due to the neutral Higgs bosons (NHBs) exchange diagrams, whose forms and the corresponding Wilson coefficients $C_{Qi}({\mu })$ can be found in [28].

As we noted earlier, in the 2HDM of type III, both the doublets can couple to the up-type and down-type quarks, and without loss of generality, we can use a basis such that, the first doublet produces the masses of all the gauge-bosons and fermions in the SM, whereas all the new Higgs fields
originate from the second doublet. This choice permits us to write the flavor changing (FC) part of the Yukawa Lagrangian at tree level as:
\begin{eqnarray}
{\cal{L}}^{III}_{Y,FC}=
\xi^{U}_{ij} \bar{Q}_{i L} \tilde{\phi_{2}} U_{j R}+
\xi^{D}_{ij} \bar{Q}_{i L} \phi_{2} D_{j R} +\xi^{\ell}_{ij} \bar{\ell}_{i L} \phi_{2} \ell_{j R}+ h.c.,
\end{eqnarray}
where $i$, $j$ are the generation indices, $\tilde{\phi_{2}}= i \sigma_2 \phi_{2}$. $\xi_{ij}^{U,D,\ell}$ are in general a non-diagonal coupling matrices, $Q_{iL}$ is the left-handed fermion
doublet, $U_{i R}$ and $D_{j R}$ are the right-handed singlets, and $\bar{\ell}_{i L}$ and $\ell_{j R}$ represent left handed SU(2) lepton doublet, right-handed SU(2) singlet, respectively. In equation (2) all states are weak states, and can be transformed to the mass eigenstates by
rotation.

After performing a proper rotation and diagonalization of the mass matrices for fermions and for Higgses, the flavor changing part of the Yukawa Lagrangian is re-expressed in terms of mass eigenstates as
follows [29]:
\begin{eqnarray}
{\cal L}^{III}_{Y,FC} &=& \frac{1}{\sqrt{2}}\left[\bar U_{i}\hat\xi^{U}_{ij}U_{j}+\bar D_{i}\hat\xi^{D}_{ij}D_{j}+\bar \ell_{i}\hat \xi^{\ell}_{ij}\ell_{j}\right]H^{0}\nonumber \\
 &-&\frac{i}{\sqrt{2}}\left[\bar U_{i}\gamma _{5} \hat\xi^{U}_{ij}U_{j}+\bar D_{i}\gamma _{5}\hat \xi^{D}_{ij}D_{j}+\bar \ell_{i}\gamma _{5} \hat\xi^{\ell}_{ij}\ell_{j}\right]A^{0}\nonumber \\
 &+& \bar U_{i}\left[\hat \xi^{U}_{ij} V_{CKM}~P_{L} -V_{CKM}~ \hat \xi^{D}_{ij}P_{R} \right]D_{j}~H^{+}+ h.c.,
\end{eqnarray}
where $U_{i}$, $D_{i}$,  $\ell_{i}$ are mass eigenstates of up- and down-type quarks and leptons, $H^{0}$, $A^{0}$ are CP- even and
-odd neutral Higgses, and $H^{\pm}$ are charged Higgses. $\xi^{U, D, \ell}_{ij}$ are the FC Yukawa quark matrices for mass eigenstates which include all the FCNC couplings. $V_{CKM}=(V_{L}^{U}){^\dag}V_{L}^{D}$ is the usual Cabibbo-Kobayashi-Maskawa (CKM) matrix, and $P_{L(R)}=\frac{(1\mp \gamma)_{5}}{2}$ are the projection operators. Because the definition of the $\hat\xi^{U,D,\ell}_{ij}$ couplings is arbitrary, in order to proceed further, in this work we adopt the Cheng-Sher ansatz [29],
\begin{eqnarray}
\hat \xi^{U,D,\ell}_{ij} = \lambda_{ij}\frac{\sqrt{2m_i m_j}}{v},
\end{eqnarray}
where, $v$ is the SM vacuum expectation value (vev), $v = 246$ GeV. This ansatz ensures that the FCNC within the first two generations are naturally suppressed by small quark masses.

In essence, from equation (3), it is clear that the $b\rightarrow  s\ell ^{+}\ell ^{-}$ transition at tree level receives contributions by exchanging neutral $H^{0}$ and $A^{0}$ Higgs bosons diagrams, like,
$\sim  \hat\xi_{bs}~\hat\xi_{\mu\mu}\frac{1}{q^2-M^2_{H^{0}(A^{0})}}\left[(\bar b(\gamma _{5})s)(\bar \mu (\gamma _{5})\mu)\right]H^{0}(A^{0})$. In the following, we assume that the neutral Higgs bosons masses are heavy enough to avoid such contributions to $B_s \to \mu^+\mu^-$ decay and similar related processes [5], and references therein. Thus, from
the above discussion it is clear that, we can safely neglect the NHBs exchange diagrams, and the transition $b \rightarrow  s \ell^+ \ell^-$ receives only contributions at loop level by exchanging the $W^{\pm}$, $Z$, $\gamma$ and charged Higgs boson fields. Interestingly, the charged Higgs boson exchange diagrams do not produce new operators for the $b\rightarrow s\ell ^{+}\ell ^{-}$ transition, and only modify the value of the SM Wilson coefficients [30].

Consequently, the operators responsible for the dileptonic $\Lambda_{b}\to \Lambda\ell^{+}\ell^{-}$ decay are only $O_{7}$, $O_{9}$, $O_{10}$, and the corresponding effective Hamiltonian in the SM and 2HDM for the $b\rightarrow s\ell^{+}\ell^{-}$ transition can be written as:
\begin{eqnarray}
H_{eff}(b \rightarrow s\ell ^{+}\ell ^{-})&=&\frac{G_{F}\alpha_{em} }{2\sqrt{2}\pi }V_{tb}V_{ts}^{\ast }\bigg\{C_{9}^{2HDM}(\mu )\bar{s}\gamma _{\mu }(1- \gamma_5)b(\bar{\ell}\gamma ^{\mu }\ell) \nonumber \\
&+&C_{10}^{2HDM}(\mu )\bar{s}\gamma _{\mu }(1- \gamma_5)b(\bar{\ell}\gamma ^{\mu }\gamma _{5}\ell)\nonumber \\
&-&2m_{b}C_{7}^{2HDM}(\mu )\bar{s}i\sigma _{\mu \nu }\frac{q^{\nu }}{q^{2}}(1+ \gamma_5)b(\bar{\ell}\gamma ^{\mu }\ell)\bigg\},
\end{eqnarray}
with [30]:
\begin{eqnarray}
C_{7}^{2HDM}(\mu)&=&C_{7}^{SM}(\mu)+\left|\lambda_{tt}\right|^2\left(\frac{y(7-5y-8y^2)}{72(y-1)^3}+\frac{y^2(3y-2)}{12(y-1)^4}\mathrm{ln}y\right) \nonumber \\
&&+\left|\lambda_{tt}\lambda_{bb}\right|e^{i\theta}\left(\frac{y(3-5y)}{12(y-1)^2}+\frac{y(3y-2)}{6(y-1)^3}\mathrm{ln}y\right),  \\
C_{9}^{2HDM}(\mu)&=&C_{9}^{SM}(\mu)+\left|\lambda_{tt}\right|^2\bigg[\frac{1-4sin^2\theta_W}{sin^2\theta_W}\frac{xy}{8}\left(\frac{1}{y-1}-\frac{1}{(y-1)^2}\mathrm{ln}y\right) \nonumber \\
&&-y\left(\frac{47y^2-79y+38}{108(y-1)^3}-\frac{3y^3-6y+4}{18(y-1)^4}\mathrm{ln}y\right)\bigg], \\
C_{10}^{2HDM}(\mu)&=&C_{10}^{SM}(\mu)+\left|\lambda_{tt}\right|^2\frac{1}{sin^2\theta_W}\frac{xy}{8}\left(-\frac{1}{y-1}+\frac{1}{(y-1)^2}\mathrm{ln}y\right),
\end{eqnarray}
where $x=\frac{m^2_t}{m^2_{W}}$, and $y=\frac{m^2_t}{m^2_{H^\pm}}$. The calculations of the Wilson coefficients $C^{SM}_{7}({\mu })$,  $C^{SM}_{9}({\mu })$, and  $C^{SM}_{10}({\mu })$ are performed at Next-to-Leading Order (NLO), and their explicit expressions at the energy scale $\mu=m_{b}$ are given in [31], while their numerical values are listed in Table 1.

In the 2HDM, the free parameters are the mass of the charged Higgs boson $m_{H^{\pm}}$, and the coefficients $\lambda_{tt}$, $\lambda_{bb}$. The coefficients $\lambda_{tt}$ and $\lambda_{bb}$ for type III of 2HDM are complex parameters of order ${\cal O}(1)$, so that $\lambda_{tt}\lambda_{bb} \equiv |\lambda_{tt}\lambda_{bb} | e^{i \theta}$, where $\theta$ is the only single CP phase of the vacuum in this version. In this way, $\lambda_{ij}$ allow the charged Higgs boson to interfere destructively or constructively to the SM contributions.

Additionally, the Wilson coefficient $C_{9}^{2HDM}(\mu)$ receives long distance contributions coming from the charmonium resonances $J/\psi$, $\psi^\prime$, $\cdots$, which is replaced by an effective coefficient $C_{9}^{eff}(\mu )$ [31]:
\begin{equation}
C_{9}^{eff}(\mu )=C_{9}^{2HDM}(\mu )~\eta(s^{\prime })+Y_{SD}(n,s^{\prime})+Y_{LD}(n,s^{\prime }),
\end{equation}%
where the parameters $n$ and $s^{\prime }$ are defined as $n=m_{c}/m_{b}$, $s^{\prime}=q^{2}/m_{b}^{2}$, whereas, $\eta(s^{\prime })=1 +\frac{\alpha_s(\mu)}{\pi}\, \omega(s^{\prime })$, with
\begin{eqnarray}
\omega(s^{\prime}) & = & - \frac{2}{9}
\pi^2 - \frac{4}{3}\mbox{Li}_2(s^{\prime}) - \frac{2}{3}
\ln s^{\prime} \ln(1-s^{\prime}) - \frac{5+4s^{\prime}}{3(1+2s^{\prime})}\ln(1-s^{\prime}) \nonumber \\
&& - \frac{2 s^{\prime} (1+s^{\prime}) (1-2s^{\prime})}{3(1-s^{\prime})^2
(1+2s^{\prime})} \ln s^{\prime} + \frac{5+9s^{\prime}-6{s^{\prime}}^2}{6 (1-s^{\prime}) (1+2s^{\prime})},
\end{eqnarray}

\begin{eqnarray}
Y_{SD}(n,s^{\prime }) &=&h(n,s^{\prime })\left[3C_{1}(\mu )+C_{2}(\mu
)+3C_{3}(\mu )+C_{4}(\mu )+3C_{5}(\mu )+C_{6}(\mu )\right]  \notag \\
&&-\frac{1}{2}h(1,s^{\prime })\left[4C_{3}(\mu )+4C_{4}(\mu )+3C_{5}(\mu)+C_{6}(\mu )\right]-\frac{1}{2}h(0,s^{\prime })\left[C_{3}(\mu )+3C_{4}(\mu )\right]  \notag \\
&&+{\frac{2}{9}}\left[3C_{3}(\mu )+C_{4}(\mu )+3C_{5}(\mu )+C_{6}(\mu )\right],
\end{eqnarray}
\begin{eqnarray}
Y_{LD}(n,s^{\prime }) &=&\frac{3}{\alpha _{em}^{2}}\left[3C_{1}(\mu )+C_{2}(\mu
)+3C_{3}(\mu )+C_{4}(\mu )+3C_{5}(\mu )+C_{6}(\mu )\right]  \notag \\
&&\times\sum_{j=\psi ,\psi ^{\prime }}k_{j}\frac{\pi \Gamma
(j\rightarrow l^{+}l^{-})M_{j}}{q^2-M_{j}^{2}+iM_{j}\Gamma _{j}^{tot}},
\label{LD}
\end{eqnarray}
where
\begin{eqnarray}
h(n,s^{\prime }) &=&-{\frac{8}{9}}\mathrm{ln}n+{\frac{8}{27}}+{\frac{4}{9}}x-%
{\frac{2}{9}}(2+x)|1-x|^{1/2}\left\{
\begin{array}{l}
\ln \left\vert \frac{\sqrt{1-x}+1}{\sqrt{1-x}-1}\right\vert -i\pi \quad
\mathrm{for}{{\ }x\equiv 4n^{2}/s^{\prime }<1} \\
2\arctan \frac{1}{\sqrt{x-1}}\qquad \mathrm{for}{{\ }x\equiv
4n^{2}/s^{\prime }>1}%
\end{array}%
\right. ,  \notag \\
h(0,s^{\prime }) &=&{\frac{8}{27}}-{\frac{8}{9}}\mathrm{ln}{\frac{m_{b}}{\mu
}}-{\frac{4}{9}}\mathrm{ln}s^{\prime }+{\frac{4}{9}}i\pi,
\end{eqnarray}
and $k_{j}$ are phenomenological parameters introduced to compensate for vector meson dominance and the factorization approximation, and are taken to be $\kappa \cong 1.0$ and $\kappa\cong 2.0$ for the lowest
resonances $J/\psi$ and $\psi^{\prime}$ respectively,
$M_{j}$, $\Gamma _{j}^{tot}$, and $\Gamma (j\rightarrow l^{+}l^{-})$ are the masses, total widths, and partial widths of the resonances, respectively.

Further, the Wilson coefficient $C_{7}^{2HDM}(\mu)$ receives another non-factorizable effects coming from the charm loop which can bring
further corrections to the radiative $b\rightarrow s\gamma $ transition [32]:
\begin{equation}
C_{7}^{eff}(\mu )=C_{7}^{2HDM}(\mu )+C_{b\rightarrow s\gamma }(\mu ),
\label{c7coefficient}
\end{equation}
while, the Wilson coefficient $C_{10}^{2HDM}$ does not change under the renormalization procedure mentioned above, and so it is independent of the energy scale, so we will let $ C_{10}^{eff}(\mu) \equiv C_{10}^{2HDM}(m_W)$.

In terms of the above criteria, the effective Hamiltonian of equation (5) in the 2HDM can be re-written as:
\begin{eqnarray}
H_{eff}(b \rightarrow s\ell ^{+}\ell ^{-})&=&\frac{G_{F}\alpha_{em}}{2\sqrt{2}\pi }V_{tb}V_{ts}^{\ast }\bigg\{C_{9}^{eff}(\mu )\bar{s}\gamma _{\mu }(1- \gamma_5)b(\bar{\ell}\gamma ^{\mu }\ell) \nonumber \\
&+&C_{10}^{eff}(\mu )\bar{s}\gamma _{\mu }(1- \gamma_5)b(\bar{\ell}\gamma ^{\mu }\gamma _{5}\ell) \nonumber \\
&-&2m_{b}C_{7}^{eff}(\mu )\bar{s}i\sigma _{\mu \nu }\frac{q^{\nu }}{q^{2}}(1+ \gamma_5)b(\bar{\ell}\gamma ^{\mu }\ell)\bigg\},
\end{eqnarray}
where $q$ is the sum of 4 momenta of $\ell ^{+}$ and $\ell ^{-}$, $\alpha_{em}$
is the fine structure constant, $C_7^{eff}(\mu )$, $C_9^{eff}(\mu )$, and $C_{10}^{eff}(\mu )$ are renamed to be the
modified Wilson coefficients representing the different interactions including both the SM and 2HDM contributions.

\section{{\small Phenomenological observables of $\Lambda_{b}\to \Lambda\ell^{+}\ell^{-}$ decay in the 2HDM}}

\subsection{{\small Matrix elements}}
To get the matrix elements for $\Lambda_{b}\to \Lambda\ell^{+}\ell^{-}$ decay, it is necessary to sandwich the effective
Hamiltonian of $b\rightarrow s\ell ^{+}\ell ^{-}$ between the initial and final baryon states. Equation (15) has four basic hadronic matrix elements $\langle \Lambda |\bar s \gamma_\mu b | \Lambda_{b}\rangle $, $\langle \Lambda |\bar s \gamma_\mu \gamma_5 b |  \Lambda_{b}\rangle$, $\langle\Lambda |\bar s i \sigma_{\mu \nu} q^\nu b |  \Lambda_{b} \rangle$, and $\langle \Lambda |\bar s i \sigma_{\mu \nu}\gamma_5 q^\nu b |  \Lambda_{b} \rangle$.  These hadronic matrix elements in terms of a set of unknown form factors are parameterized as [33]:
\begin{eqnarray}
\langle \Lambda |\bar s \gamma_\mu b | \Lambda_{b}\rangle & =& \bar u_\Lambda \Big[
f_1(q^2) \gamma_\mu + i f_2(q^2) \sigma_{\mu \nu} q^\nu +f_3(q^2) q_\mu \Big]
u_{ \Lambda_{b}}\;,\nonumber\\
\langle \Lambda |\bar s \gamma_\mu \gamma_5 b |  \Lambda_{b}\rangle & =&
\bar u_\Lambda \Big[
g_1(q^2) \gamma_\mu \gamma_5 + i g_2(q^2) \sigma_{\mu \nu}\gamma_5 q^\nu +
g_3(q^2) \gamma_5 q_\mu \Big]
u_{ \Lambda_{b}}\;,\nonumber\\
\langle\Lambda |\bar s i \sigma_{\mu \nu} q^\nu b |  \Lambda_{b} \rangle & =&
\bar u_\Lambda \Big[
f_1^T(q^2) \gamma_\mu + i f_2^T(q^2) \sigma_{\mu \nu} q^\nu +f_3^T(q^2) q_\mu \Big]
u_{ \Lambda_{b}}\;,\nonumber\\
\langle \Lambda |\bar s i \sigma_{\mu \nu}\gamma_5 q^\nu b |  \Lambda_{b} \rangle & =&
\bar u_\Lambda \Big[
g_1^T(q^2) \gamma_\mu \gamma_5 + i g_2^T(q^2) \sigma_{\mu \nu}\gamma_5 q^\nu +
g_3^T(q^2) \gamma_5 q_\mu \Big]
u_{ \Lambda_{b}}\;.
\end{eqnarray}

Currently, there are some studies in the literature on $\Lambda_{b}\to \Lambda$ transition form factors
such as heavy quark effective theory (HQET) [34], lattice QCD calculations [35], light-cone sum rules approach [36], perturbative QCD approach [37], QCD sum rule approach [38], and Bethe-Salpter equation approach [39]. In the present work, based on lattice QCD calculations [35], the form factors
$f_{1}(q^2)$, $f_{2}(q^2)$, $f_{3}(q^2)$, $g_{1}(q^2)$, $g_{2}(q^2)$, $g_{3}(q^2)$, $f_{1}^T(q^2)$, $f_{2}^T(q^2)$, $f_{3}^T(q^2)$, $g_{1}^T(q^2)$, $g_{2}^T(q^2)$ and $g_{3}^T(q^2)$ introduced above in equation (16) are related to the helicity form factors of [35] as follows:

\begin{eqnarray}
 f_+(q^2)     &=& f_1(q^2) - \frac{q^2}{(m_{\Lambda_b}+m_\Lambda)} f_2(q^2),  \\
 f_\perp(q^2) &=& f_1(q^2) - (m_{\Lambda_b}+m_\Lambda)f_2(q^2),  \\
 f_0(q^2)     &=& f_1(q^2) + \frac{q^2}{(m_{\Lambda_b}-m_\Lambda)} f_3(q^2), \\
 g_+(q^2)     &=& g_1(q^2) +\frac{q^2}{(m_{\Lambda_b}-m_\Lambda)}g_2(q^2), \\
 g_\perp(q^2) &=& g_1(q^2) + (m_{\Lambda_b}-m_\Lambda)g_2(q^2), \\
 g_0(q^2)     &=& g_1(q^2) - \frac{q^2}{(m_{\Lambda_b}+m_\Lambda)} g_3(q^2), \\
 h_+(q^2)     &=& f_2^{T}(q^2)+ \frac{(m_{\Lambda_b}+m_\Lambda)}{(m_{\Lambda_b}-m_\Lambda)}f_{3}^T(q^2),  \\
 h_\perp(q^2) &=& f_2^{T}(q^2) - \frac{f_1^{T}(q^2)}{(m_{\Lambda_b}+m_\Lambda)},  \\
 \widetilde{h}_+(q^2)     &=& g_2^{T}(q^2) - g_3^{T}(q^2),  \\
 \widetilde{h}_\perp(q^2) &=& g_2^{T}(q^2) + \frac{g_1^{T}(q^2)}{(m_{\Lambda_b}-m_\Lambda)}.
\end{eqnarray}

The above helicity form factors $f_+(q^2)$, $ f_\perp(q^2)$, $f_0(q^2)$,  $ g_+(q^2) $, $ g_\perp(q^2)$, $g_0(q^2)$, $ h_+(q^2)$, $h_\perp(q^2)$, $\widetilde{h}_+(q^2)$, and $ \widetilde{h}_\perp(q^2) $ are parameterized based on lattice QCD calculations in the following way [35]:
\begin{equation}
f(q^2) = \frac{1}{1-q^2/(m_{\rm pole}^f)^2} \big[ a_0^f + a_1^f\:\kappa(q^2) \big],
\end{equation}
where,
\begin{equation}
\kappa(q^2) = \frac{\sqrt{t_+-q^2}-\sqrt{t_+-t_0}}{\sqrt{t_+-q^2}+\sqrt{t_+-t_0}}.
\end{equation}
Here, $t_0 = q^2_{\rm max} = (m_{\Lambda_b} - m_{\Lambda})^2$, $t_+ = (m_B + m_K)^2$, $m_B=5.279~GeV$, $m_K=0.494~GeV$, and values of the parameters $a_0^f$, $a_1^f$, and $m_{\rm pole}^f$ to the first-order fit are collected in Table 2.

With the above definitions of transition matrix elements, and the form factors, we get the effective amplitude for $\Lambda _{b}\rightarrow \Lambda \ell^{+}\ell^{-}$ decay:
\begin{eqnarray}
\mathcal{M}(\Lambda _{b}\rightarrow \Lambda \ell^{+}\ell^{-}) &=& {G_F \alpha_{em} \over 4 \sqrt{2} \pi} V_{tb} V_{ts}^\ast \Bigg\{
\bar{\ell} \gamma^\mu \ell \, \bar{u}_\Lambda
\Big[ A_1 \gamma_\mu (1+\gamma_5) + B_1 \gamma_\mu (1-\gamma_5) \nonumber\\
&+& i \sigma_{\mu\nu} q^\nu \Big( A_2 (1+\gamma_5)+ B_2 (1-\gamma_5) \Big)\nonumber\\
&+& q_\mu \Big(  A_3 (1+\gamma_5) + B_3 (1-\gamma_5) \Big) \Big]u_{\Lambda_b}  \nonumber\\
&+& \bar{\ell} \gamma^\mu \gamma_5 \ell \, \bar{u}_\Lambda
\Big[ D_1 \gamma_\mu (1+\gamma_5) + E_1 \gamma_\mu (1-\gamma_5) \nonumber\\
&+& i \sigma_{\mu\nu} q^\nu \Big( D_2 (1+\gamma_5) +E_2 (1-\gamma_5) \Big) \nonumber\\
&+& q_\mu \Big(  D_3 (1+\gamma_5) + E_3 (1-\gamma_5) \Big) \Big]u_{\Lambda_b}  \Bigg\},
\end{eqnarray}
where the various functions $A_i,~B_i$, $D_j$, and $E_j$ ($i,j=1,2,3$) are defined as:
\begin{eqnarray}
A_i &=& C_9^{eff}(f_i-g_i)-\frac{2m_bC_7^{eff} }{q^2}
(f_i^T+g_i^T)\;,\nonumber\\
B_i &=& C_9^{eff}(f_i+g_i)-\frac{2m_bC_7^{eff}}{q^2}
(f_i^T-g_i^T)\;,\nonumber\\
D_j &=&C_{10}^{eff} (f_j-g_j)\;,~~~~
E_j = C_{10}^{eff}(f_j+g_j)\;.
\end{eqnarray}

\subsection{{\small The differential decay rate for $\Lambda_{b}\to \Lambda\ell^{+}\ell^{-}$}}

The differential decay rate of $\Lambda_{b}\to \Lambda\ell^{+}\ell^{-}$ is given by:
\begin{eqnarray}
d\Gamma=\frac{1}{4m_{\Lambda_b}}\left(\prod_{f}\frac{d^3p_{f}}{(2\pi)^3}\frac{1}{2E_{f}}\right)(2\pi)^4\delta^{4}\left(p_{\Lambda_b}-\sum_{f}p_{f}\right)\mathcal{|\overline{M}|}^{2}.
\end{eqnarray}
Here, $\mathcal{|\overline{M}|}^{2}$ is the squared amplitude averaged over the initial
polarization and summed over the final polarizations. After lengthy, but straightforward calculations, one can get the double differential decay rate in terms of the various form factors in the 2HDM:
\begin{eqnarray}
\frac{d^2 \Gamma (\hat s,z)}{d\hat s dz}=\frac{G_F^2~ \alpha^2_{em}}{2^{14}\pi^5}~
|V_{tb} V_{ts}^*|^2~~m_{\Lambda_{b}} v_\ell~ \lambda^{1/2}(1, r, \hat s)~
\mathrm{T}(\hat s ,z)\;,
\end{eqnarray}
where $\hat s=q^2/m_{\Lambda_b}^2$, $z=\cos \theta $ is the angle
between $p_{\Lambda_b}$ and  $p_{+}$  in the center of mass
frame of $\ell^+ \ell^-$ pair, $v_\ell=\sqrt{1-4m_{\Lambda_b} m_\ell^2/ \hat s}$,
$\lambda(1,r,\hat{s}) = 1 + r^2 + \hat{s}^2 - 2 r
- 2 \hat{s} - 2 r \hat{s}$, is the usual
triangle function. The function $\mathrm{T}(\hat s ,z)$ is given by:
\begin{eqnarray}
\mathrm{T}(\hat s ,z)=\mathrm{T_{0}}(\hat s)+z~ \mathrm{T_{1}}(\hat s)+z^2~\mathrm{T_{2}}(\hat s),
\end{eqnarray}
with
\begin{eqnarray}
\mathrm{T_{0}}(\hat s) &=& 32 m_\ell^2 m_{\Lambda_b}^2\hat s(1+r -\hat s)(|D_3|^2+|E_3|^2)\nonumber\\
&+&
64 m_\ell^2 m_{\Lambda_b}^3(1-r -\hat s)Re(D_1^*E_3+D_3 E_1^*)
+64  m_{\Lambda_b}^2 \sqrt{r} (6 m_\ell^2-\hat s m_{\Lambda_b}^2)Re(D_1^*E_1)
\nonumber\\
&+&64 m_\ell^2 m_{\Lambda_b}^3 \sqrt{r} \Big(
2 m_{\Lambda_b} \hat s Re(D_3^*E_3)+(1-r+\hat s)Re(D_1^* D_3+ E_1^* E_3)\Big)\nonumber\\
&+&32 m_{\Lambda_b}^2 (2 m_\ell^2+m_{\Lambda_b}^2 \hat s)\Big((1-r +\hat s)m_{\Lambda_b}
\sqrt{r}Re(A_1^*A_2+B_1^* B_2)\nonumber\\
&-& m_{\Lambda_b}(1-r-\hat s) Re(A_1^* B_2+A_2^* B_1)-2 \sqrt{r}\Big[Re(A_1^* B_1)
+m_{\Lambda_b}^2\hat s Re(A_2^* B_2)\Big]\Big)\nonumber\\
&+& 8 m_{\Lambda_b}^2\Big(4 m_\ell^2(1+r-\hat s)+m_{\Lambda_b}^2[(1-r)^2-\hat s^2]\Big)
\Big(|A_1|^2+|B_1|^2\Big)\nonumber\\
&+& 8 m_{\Lambda_b}^4\Big(4 m_\ell^2[\lambda+(1+r-\hat s)\hat s]+m_{\Lambda_b}^2
\hat s[(1-r)^2-\hat s^2]\Big)
\Big(|A_2|^2+|B_2|^2\Big)\nonumber\\
&-& 8 m_{\Lambda_b}^2\Big(4 m_\ell^2(1+r-\hat s)-m_{\Lambda_b}^2[(1-r)^2-\hat s^2]\Big)
\Big(|D_1|^2+|E_1|^2\Big)\nonumber\\
&+& 8 m_{\Lambda_b}^5 \hat s  v_\ell^2 \Big(-8 m_{\Lambda_b} \hat s \sqrt{r}
Re(D_2^* E_2)+4 (1-r+\hat s)\sqrt{r}Re(D_1^* D_2+E_1^* E_2)\nonumber\\
&-&4(1-r -\hat s) Re(D_1^* E_2+D_2^* E_1)
+m_{\Lambda_b}[(1-r)^2-\hat s^2]
\Big[|D_2|^2+|E_2|^2\Big]\Big),
\end{eqnarray}
\begin{eqnarray}
\mathrm{T_{1}}(\hat s) &=& -16  m_{\Lambda_b}^4\hat s v_\ell \sqrt{\lambda(1,r,\hat{s})}
\Big\{ 2 Re(A_1^* D_1)-2Re(B_1^* E_1)\nonumber\\
&+& 2m_{\Lambda_b}
Re(B_1^* D_2-B_2^* D_1+A_2^* E_1-A_1^*E_2)\Big\}\nonumber\\
&+&32 m_{\Lambda_b}^5 \hat s~ v_\ell \sqrt{\lambda(1,r,\hat{s})} \Big\{
m_{\Lambda_b} (1-r)Re(A_2^* D_2 -B_2^* E_2)\nonumber\\
&+&
\sqrt{r} Re(A_2^* D_1+A_1^* D_2-B_2^*E_1-B_1^* E_2)\Big\},
\end{eqnarray}
and
\begin{eqnarray}
\mathrm{T_{2}}(\hat s)&= & 8m_{\Lambda_b}^6 v_\ell^2~ \lambda(1,r,\hat{s}) \hat s~ \Big (
(|A_2|^2+|B_2|^2+|D_2|^2+|E_2|^2\Big)\nonumber\\
&-&8m_{\Lambda_b}^4 v_\ell^2 ~\lambda(1,r,\hat{s})~\Big(|A_1|^2+|B_1|^2+|D_1|^2+|E_1|^2\Big).
\end{eqnarray}
The unpolarized differential decay rate of $\Lambda_{b}\to \Lambda\ell^{+}\ell^{-}$ can be obtained from equation (32) by integrating
out the angular dependent variable $z$ which, in turn yields:
\begin{eqnarray}
\left (\frac{d \Gamma (\hat s)}{d \hat s}\right )_0= \frac{G_F^2~ \alpha^2_{em} m_{\Lambda_b}}
{2^{13} \pi^5}~
|V_{tb}V_{ts}^*|^2 v_\ell~ \lambda^{1/2}(1, r, \hat s)~\Big[\mathrm{T_{0}}(\hat s)+ \frac{1}{3}\mathrm{T_{2}}(\hat s)\Big].
\end{eqnarray}

\subsection{{\small  Leptons forward-backward asymmetry of $\Lambda_{b}\to \Lambda\ell^{+}\ell^{-}$}}

Another useful observable to look for new physics effects in $\Lambda_{b}\to \Lambda\ell^{+}\ell^{-}$ decay is the leptons forward-backward asymmetry ($A_{FB}$). Since $A_{FB}$ depends
on the chirality of the hadronic and leptonic currents, therefore, this observable is very sensitive to new physics beyond the SM through shifting its zero value position. To calculate the leptons forward-backward asymmetry, we consider the double
differential decay rate formula for the process $\Lambda_{b}\to \Lambda\ell^{+}\ell^{-}$ defined in equation (32).

The normalized leptons forward-backward asymmetry
is defined as:
\begin{eqnarray}
A_{FB}(\hat s)=\frac{\int_0^1 \frac{d \Gamma}{d \hat s dz}dz-\int_{-1}^0
\frac{d \Gamma}{d \hat s dz}dz}
{\int_0^1 \frac{d \Gamma}{d \hat s dz}dz+\int_{-1}^0
\frac{d \Gamma}{d \hat s dz}dz}.
\end{eqnarray}

Following the same procedure as we did for the differential decay rate, one
can easily get the expression for the leptons forward-backward asymmetry:
\begin{eqnarray}
A_{FB}(\hat s)=\frac{\mathrm{T_{1}}(\hat s)}{2\left(\mathrm{T_{0}}(\hat s)+\mathrm{T_{2}}(\hat s )/3\right)}.
\end{eqnarray}
\subsection{{\small  $\Lambda_{b}\to \Lambda\ell^{+}\ell^{-}$ decay with polarized $\Lambda$}}

Now let us consider the case when the final $\Lambda$ baryon is polarized. As we already noted, unlike mesonic decays, the baryonic decays could keep the helicity structure of
their interactions. The importance of the polarization of baryons in general is coming in due to their right-handed couplings which are suppressed
in the standard model, and they include different combinations of structures within the  Wilson coefficients $C^{eff}_{7}(\mu)$, $C^{eff}_{9}(\mu)$, and $C^{eff}_{10}(\mu)$. For this reason, the baryonic decays are considered to be one of the promising tools to search for new physics beyond the standard model.

Here, we present the formulas for the polarized differential decay rates of $\Lambda_{b}\to \Lambda\ell^{+}\ell^{-}$. In the calculations, we
have included the lepton masses, and we define a four-dimensional spin vector for the polarized $\Lambda$ baryon in terms of a unit vector, $\hat \xi$, along the direction of $\Lambda$ spin in its rest frame as:
\begin{eqnarray}
s_\mu= \Big(\frac{\overrightarrow{p_\Lambda} \cdot \hat \xi}{m_\Lambda}\;,~~ \hat \xi+
\frac{\overrightarrow{p_\Lambda} \cdot \hat \xi}{m_\Lambda(E_\Lambda+m_\Lambda)}\overrightarrow{p_\Lambda} \Big).
\end{eqnarray}

We also introduce three orthogonal unit vectors along the
longitudinal, transverse and normal components of $\Lambda$ polarization in the $\Lambda_{b}$ rest frame as:
\begin{eqnarray}
\hat e_L= \frac{\overrightarrow{p_\Lambda}}{|\overrightarrow{p_\Lambda}|}\;,~~~
\hat e_T= \frac{\overrightarrow{p_+} \times \overrightarrow{p_\Lambda}}{|\overrightarrow{p_+} \times
\overrightarrow{p_\Lambda}|}\;,~~~~\hat e_N=\hat e_T \times \hat e_L,
\end{eqnarray}
where $\overrightarrow{p_\Lambda}$ and $\overrightarrow{p_+}$ are three-dimensional vector momenta of the $\Lambda$ and $\ell^+$
in the center of mass of the $\ell^+ \ell^-$ system. With these spin vectors, one can
get the polarized differential decay rate for any spin direction $\hat \xi $
along the $\Lambda$ baryon spin components:
\begin{eqnarray}
\frac{d \Gamma(\hat \xi)}{d \hat s}=
\frac{1}{2}\left (\frac{d \Gamma}{d \hat s} \right )_0\Big[
1+\left ( P_L ~\hat e_L + P_N ~\hat e_N +P_T~ \hat e_T \right )\cdot \hat \xi
\Big],
\end{eqnarray}
where $P_L$, $P_N$ and $P_T$ are the
longitudinal, normal and transverse polarizations of $\Lambda$ baryon, respectively, and $\left (d \Gamma/d \hat s \right )_0$ is the unpolarized decay width defined in equation (37). These polarizations asymmetries $P_i$ ($i=L, N, T$) are obtained from:
\begin{eqnarray}
P_i(\hat s)= \frac{\frac{d \Gamma}{d\hat s}(\hat \xi= \hat e_i)-\frac{d \Gamma}{d\hat s}(\hat \xi= -\hat e_i)}{\frac{d \Gamma}{d\hat s}(\hat \xi= \hat e_i)+\frac{d \Gamma}{d\hat s}(\hat \xi= -\hat e_i)},
\end{eqnarray}
where
\begin{eqnarray}
P_L(\hat s) &=& \frac{16 m_{\Lambda_{b}}^2 \sqrt{\lambda(1,r,\hat{s})}}{
2(\mathrm{T_{0}}(\hat s)+\mathrm{T_{2}}(\hat s )/3)}\biggr[ 8 m_\ell^2 m_{\Lambda_{b}}\Big
( Re(D_1^* E_3 -D_3^* E_1)+
\sqrt{r} Re(D_1^* D_3 -E_1^* E_3) \Big) \nonumber\\
&-& 4 m_\ell^2 m_{\Lambda_{b}}^2 \hat s \Big(|D_3|^2-|E_3|^2
\Big) -4 m_{\Lambda_{b}}(2 m_\ell^2+m_{\Lambda_{b}}^2
\hat s)Re(A_1^* B_2-A_2^* B_1)\nonumber\\
&-& \frac{4}{3} m_{\Lambda_{b}}^3 \hat s~ v_\ell^2\Big(3 Re(D_1^* E_2-D_2^* E_1)
+\sqrt{r} Re(D_1^* D_2-E_1^* E_2)\Big)\nonumber\\
&-& \frac{4}{3}m_{\Lambda_{b}} \sqrt{r} (6 m_\ell^2 +m_{\Lambda_{b}}^2 \hat s~ v_\ell^2)
Re(A_1^* A_2-B_1^* B_2)-\frac{2}{3} m_{\Lambda_{b}}^4 \hat s (2-2r+\hat s)
v_\ell^2 (|D_2|^2-|E_2|^2)\nonumber\\
&+& (4 m_\ell^2+m_{\Lambda_{b}}^2(1-r+\hat s))(|A_1|^2-|B_1|^2)
-(4m_\ell^2-m_{\Lambda_{b}}^2(1-r+\hat s))(|D_1|^2-|E_1|^2)\nonumber\\
&-&\frac{1}{3} m_{\Lambda_{b}}^2(1-r-\hat s)~ v_\ell^2~ (|A_1|^2-|B_1|^2+|D_1|^2-
|E_1|^2)\nonumber\\
&-& \frac{1}{3} m_{\Lambda_{b}}^2\Big[12 m_\ell^2(1-r)+m_{\Lambda_{b}}^2\hat s(3(1-r+\hat s)
+v_\ell^2 (1-r-\hat s))\Big]\Big(|A_2|^2-|B_2|^2)\biggr],
\end{eqnarray}
\begin{eqnarray}
P_N(\hat s) &=& \frac{8 \pi m_{\Lambda_{b}}^3 v_\ell \sqrt{\hat s}}{
2(\mathrm{T_{0}}(\hat s)+\mathrm{T_{2}}(\hat s )/3)}\biggr[-2 m_{\Lambda_{b}}(1-r+\hat s)\sqrt{r}
Re(A_1^* D_1+B_1^* E_1)\nonumber\\
&+& 4 m_{\Lambda_{b}}^2 \hat s \sqrt{r} Re(A_1^* E_2+A_2^* E_1+B_1^* D_2
+B_2^* D_1)\nonumber\\
&-& 2 m_{\Lambda_{b}}^3 \hat s \sqrt{r} (1-r+\hat s)Re(A_2^* D_2+B_2^* E_2)\nonumber\\
&+& 2 m_{\Lambda_{b}} (1-r-\hat s)\Big( Re(A_1^* E_1+B_1^* D_1)+m_{\Lambda_{b}}^2
\hat s Re(A_2^* E_2+ B_2^* D_2)\Big)\nonumber\\
&-& m_{\Lambda_{b}}^2\Big((1-r)^2-\hat s^2\Big)Re(A_1^* D_2 +A_2^* D_1+B_1^*
E_2 +B_2^* E_1)
\biggr],
\end{eqnarray}
\begin{eqnarray}
P_T(\hat s) &=& - \frac{8 \pi m_{\Lambda_{b}}^3 v_l \sqrt{\hat s\lambda(1,r,\hat{s})}}{2(\mathrm{T_{0}}(\hat s)+\mathrm{T_{2}}(\hat s )/3)}\biggr[ m_{\Lambda_{b}}^2(1-r+\hat s)\Big(
Im(A_2^* D_1-A_1^* D_2)\nonumber\\
&-&Im(B_2^* E_1-B_1^* E_2)\Big)
+ 2 m_{\Lambda_{b}}\Big(Im(A_1^*E_1-B_1^* D_1)\nonumber\\
&-&m_{\Lambda_{b}}^2 \hat s
Im(A_2^* E_2-B_2^* D_2)\Big)
\biggr].
\end{eqnarray}

\section{{\small Numerical Analysis}}

In this section, we investigate the sensitivity of the branching ratio, differential branching ratio, leptons forward-backward asymmetry, and polarization asymmetries of $\Lambda$ baryon on the parameters of the 2HDM within the full kinematical interval of the dilepton invariant mass $4 m_\ell^2 \leq q^2\leq (m_{\Lambda_b}-m_\Lambda)^2$. The main arbitrary parameters of type III are $\mid \lambda_{tt}\mid $, $\mid \lambda_{bb}\mid $, $m_{H^{\pm}}$, and the phase angle $\theta$. Typically, $\mid \lambda_{tt}\mid $, and $\mid \lambda_{bb}\mid $ parameters in the Yukawa couplings of type III can be complex, $\lambda_{tt} \lambda_{bb}=\mid \lambda_{tt} \lambda_{bb}\mid e^{i\theta}$, where the range of variations for $\mid \lambda_{tt}\mid $, $\mid \lambda_{bb}\mid $ and the phase angle $\theta$ are determined from the experimental results of the electric dipole moments of neutron, $B^0 - \overline{B^0}$ mixing, $R_b\equiv \frac{\Gamma (Z\rightarrow b\overline{b})}{\Gamma (Z\rightarrow hadrons)}$, $Br(b \to s \gamma)$, and $\rho_0=\frac{M_{W}^2}{M_{Z}^2cos^2\theta}$ [39-42]. The physical regions for these parameters have become more controlled as time goes on.
The experimental bounds on the neutron electric dipole moments and $Br(b \to s \gamma)$ as well as $m_{H^{\pm}}\geq 160~GeV$ obtained at LEP II constrain $\lambda_{tt}\lambda_{bb}$  to be less than 1 and the $\theta$ to be in the range $60^{\circ} - 90^{\circ}$. Similarly, the experimental mixing parameter $x_d=\frac{\Delta M_{B}}{\Gamma _{B}}$, $\Delta M_{B}$ and $\Gamma _{B}$ being the mass difference and the average width for the $B^{0}$-meson mass eigenstates, controls  $\mid\lambda_{tt}\mid$ to be less than 0.3 [29].  Further constraints are coming from the experimental results, like $BR(B\to \tau\nu)$, $R(D^*)= \frac{\Gamma (B\rightarrow D^*\mu\nu_{\mu})}{\Gamma (B\rightarrow D^*\ell\nu_{\ell})}$, and $BR(t\to c g)$ [29]. On the other hand, the experimental value of the parameter $R_b$ constrains the size of $\mid\lambda_{bb}\mid$ to be around 50. (See for example [29, 30], and references therein). From the CLEO data of $BR(B\to X_{s}\gamma)$, some constraint on $m_{H^{\pm}}$ in model III can also be found [41].

The other main input parameters are listed in Tables 2, and 3, while the values of Wilson coefficients in the SM are presented in Table 1. The obtained numerical results are shown in Figures 1-5. From Figures it is clear that the long distance contributions (the charmonium resonances) can give
real effects on those observables by taking into account the first two low lying resonances $J/\psi$ and $\psi^{\prime}$.

For the sake of convenience, Figure 1 shows the $\mid\lambda_{tt}\mid$ dependence of the branching ratios BR($\Lambda_b \to \Lambda \ell^+\ell^-$)($\ell=%
\protect\mu, \protect\tau$) in the SM and the 2HDM of type III for $\mid\lambda_{bb}\mid =50$, $\theta =90^{\circ}$, and for different values of $m_{H^{\pm}}$ with and without LD contributions, respectively. In Figure 1, we have respected both the upper limit of $\mid\lambda_{tt}\mid \leq 0.3$ and the lower limit of $m_{H^{\pm}}\geq 160~GeV$ [1, 29, 43]. From Figure 1, it follows that, the new physics contribution of type III 2HDM can offer one maximum 6-7 times of enhancement for the branching ratio BR($\Lambda_b \to \Lambda \mu^+\mu^-$) at $m_{H^{\pm}}= 160~GeV$, and $\mid\lambda_{tt}\mid =0.3$ [22, 23]. Furthermore, one can see from figure 1 (a) that, when $0.05\leq \mid\lambda_{tt}\mid \leq 0.15$ for $160~GeV \leq m_{H^{\pm}}\leq 300~GeV$ the contribution of the type III 2HDM exceeds the SM ones by at most 1-2 times. The numerical values of the branching ratios for $\Lambda _{b}\rightarrow \Lambda \ell^{+}\ell^{-}$ ($\ell=\mu,\tau $) with and without LD contribution in SM and 2HDM are summarized in Table 4.

Generally speaking, the branching ratios in 2HDM are always exceeding the SM predictions, and become more important at $0.05\leq \mid\lambda_{tt}\mid \leq 0.15$ when $160~GeV \leq m_{H^{\pm}}\leq 300~GeV$. The general behaviour of 2HDM contributions is; an increasing in the values of $\mid\lambda_{tt}\mid$ creates an increase in the values of the branching ratios, while an increasing in the values of $m_{H^{\pm}}$, causes a decrease in the values of the branching ratios. For example, when $m_{H^{\pm}}= 700~GeV$ the branching ratios within 2HDM are exceeding the SM results slightly, but still in agreement with a recent CDF measurement BR($\Lambda_{b}\to \Lambda\mu^{+}\mu^{-}$)= $[1.73\pm 0.42(stat)\pm 0.55(syst)]\times 10^{-6}$ [22], and LHCb collaboration BR($\Lambda_{b}\to \Lambda\mu^{+}\mu^{-}$)= $[0.96\pm 0.16(stat)\pm 0.13(syst)\pm 0.21(norm)]\times 10^{-6}$ [23]. Thus, a sensitive measurement of BR($\Lambda_b \to \Lambda \ell^+\ell^-$)($\ell=\protect\mu, \protect\tau$) as well as the value of $\mid\lambda_{tt}\mid$, will play in future a critical role in establishing new physics beyond the SM, in particular 2HDM.

In Figure 2, we show the dependence of the differential branching ratios of $\Lambda_b \to \Lambda \ell^+\ell^-$($\ell=\protect\mu, \protect\tau$) on $q^2$  with and without
LD contributions by using the reference values for the parameters as specified before; $\mid\lambda_{tt}\mid =0.15$, $\mid\lambda_{bb}\mid =50$, and $\theta =90^{\circ}$. From Figure 2, the agreement of the SM with the experimental data in the dilepton invariant mass-squared region $15 < q^2 < 20~GeV^{2}$ for $\mu^{+}\mu^{-}$ channel is clear [24]. Also, one can see that the 2HDM effects are significant for both muon and tau pairs being in the final state.

In Figure 3, the leptons $A_{FB}$ for the $\Lambda_b \to \Lambda \ell^+\ell^-$ ($\ell=\protect\mu, \protect\tau$) decays as functions of $q^2$
are presented. Figure 3 (a, b) describe the leptons $A_{FB}$ in $\Lambda
_{b}\rightarrow \Lambda \mu ^{+}\mu ^{-}$ channel with and without LD contributions, from which one can easily distinguish between the SM and 2HDM. It is clear from Figure that, in the SM the zero position of $A_{FB}$ is due to the opposite sign of $C_{7}^{SM}(\mu)$ and
$C_{9}^{SM}$, whereas, in 2HDM, the sign of $C_{7}^{eff}(\mu)$ and $C_{9}^{eff}(\mu)$ are the same and have considerable contributions and hence the zero point of the $A_{FB}$ completely disappears. Moreover, it should be noted that at $q^2\leq 5~GeV^2$ the sign of $A_{FB}$ in SM and 2HDM is completely different. Thus, determining the sign of $A_{FB}$ in this domain can give unambiguous information about the existence of the charged Higgs particle. This property is considered to be one of the most promising tools in looking for new physics beyond the SM.

For $\Lambda _{b}\rightarrow \Lambda \tau ^{+}\tau ^{-}$ the $A_{FB}$ with and without LD contributions
are represented in Figure 3 (c, d). For this channel, the $A_{FB}$ are found to be insensitive to the effects coming from different $m_{H^{\pm}}$ masses in the 2HDM, and the 2HDM prediction for $A_{FB}$ in the region $q^2\simeq 15-17~ GeV^2$  is slightly smaller than the SM one.

Figure 4 shows the dependence of longitudinal polarization asymmetry of $\Lambda $
baryon on the square of momentum transfer. In Figure 4 (a, b), the effects of
2HDM show overall considerable deviations from SM results in the low momentum transfer regions $0 < q^2 < 5~GeV^{2}$ for $\Lambda_{b}\to \Lambda\mu^{+}\mu^{-}$ channel, the $P_{L}$ asymmetry of $\Lambda $ baryon is large and negative in all cases. On the other hand, for $\Lambda _{b}\rightarrow \Lambda \tau ^{+}\tau ^{-}$ channel in Figure 4 (c, d), the $P_{L}$ asymmetry of $\Lambda $ baryon in 2HDM is indistinguishable from that in the SM. Therefore, the $P_{L}$ asymmetry of $\Lambda $ baryon for the muonic mode is predictive for establishing new physics beyond the SM. Whereas, one can easily see that the
$P_{N}$  asymmetry of $\Lambda $ baryon is so sensitive to the sign of the $C_{7}^{eff}(\mu)$ in 2HDM, and its result is also distinguishable from the SM as shown
in Figure 5. The $P_{N}$  asymmetry of $\Lambda $ baryon is negative in the low momentum transfer regions $1 < q^2 < 10~GeV^{2}$, positive and large in the high momentum transfer regions $10 < q^2 < 20~GeV^{2}$. In low region, increasing $m_{H^{\pm}}$ decreases the values of the normal polarization asymmetry. For example, with $ m_{H^{\pm}}= 160~GeV$ the values of $P_{N}$  decrease of about $60\% $ than the SM results. Therefore, independent measurements of branching ratios and measurements of longitudinal and normal polarization asymmetries for $\Lambda _{b}\rightarrow \Lambda \ell^+\ell^-$ in future experiments will be a useful tool for establishing the 2HDM.

Finally, from equation (46) it is clear that the transverse polarization asymmetry of $\Lambda $ baryon is
proportional to the imaginary part of $C_{7}^{*eff}(\mu)$ and
$C_{9}^{*eff}(\mu)$. These imaginary parts are quite small in the 2HDM as well as in the SM, and hence the values of the transverse
polarization asymmetries of $\Lambda $ baryon at different values of $ m_{H^{\pm}}$ and $\mid\lambda_{tt}\mid$ are almost equal to zero. For this reason we do not show them here.

\section{Conclusion}

In this paper, we have carried out a comprehensive analysis on $\Lambda_b \to \Lambda \ell^+\ell^-$ ($\ell=\protect\mu, \protect\tau$) decays in the general 2HDM of type III.
We have calculated the branching ratios, differential branching ratios, leptons forward-backward asymmetry, and $\Lambda$ baryon polarizations using the lattice
QCD calculations of the relevant $\Lambda_b \to \Lambda $ form factors. We have shown that, the branching ratios and the differential branching ratios in 2HDM deviate sizably from that of the SM,
especially in the large momentum transfer region. For the leptons forward-backward asymmetry in $\Lambda_b \to \Lambda \ell^+\ell^-$ ($\ell=\protect\mu, \protect\tau$) decays the deviations from
the SM are very mild in 2HDM. Moreover, in the 2HDM of type III the zero-point of the $A_{FB}$ completely disappears, and the sign predicted in 2HDM is different from that in SM at $q^2\leq 5~GeV^2$ in muon channel. In short, we think that, the orders of the obtained values of branching ratios, differential branching ratios, leptons $A_{FB}$, $P_{L}$ and $P_{N}$ polarization of $\Lambda $ baryon in $\Lambda _{b}\rightarrow \Lambda \ell^{+}\ell^{-}$ ($l=\mu ,\tau $) decays can be measured at LHCb.

To conclude, even though there are several angular distributions that have been already measured at the LHCb (see for example [24]), and several updated theoretical discussion on such decay [44-46], still one needs precise measurement of such quantities, like branching ratio, differential branching ratio, lepton forward-backward asymmetry, and polarization asymmetry of $\Lambda$ baryon in $\Lambda _{b}\rightarrow \Lambda \ell^{+} \ell^{-}$ ($l=\mu, \tau$) decays which will be helpful to search for the existence of the charged Higgs particles, and open the possibility of establishing new physics beyond the SM.

\clearpage
\begin{table}[ht]
\renewcommand{\arraystretch}{1.5}
\addtolength{\arraycolsep}{3pt}
$$
\begin{array}{|c|c|c|c|c|c|c|c|c|}
\hline C_{1} & C_{2} & C_{3} & C_{4} & C_{5} & C_{6} & C_{7}^{SM} &
C_{9}^{SM} & C_{10}^{SM}\\ \hline
-0.248 & 1.107& 0.011& -0.026& 0.007& -0.031& -0.313& 4.344& -4.669\\
\hline
\end{array}
$$
\caption{The numerical values of the Wilson coefficients at $\mu =
m_{b}$ scale within the SM [31].}
\renewcommand{\arraystretch}{1}
\addtolength{\arraycolsep}{-3pt}
\end{table}

\begin{table}[ht]
\centering
\begin{tabular}{cccccc}
\hline \hline
 Parameter &  Value & $m_{\rm pole}^f$ (GeV)&  Parameter & Value&  $m_{\rm pole}$ (GeV)\\
\hline
$a_0^{f_+}$ &  0.4221 &$5.416^{*}$&  $a_1^{g_0}$ & -1.0290&$5.367^{*}$ \\
$a_1^{f_+}$ & -1.1386 &$5.416^{*}$&  $a_1^{g_\perp}$ & -1.1357&5.750 \\
$a_0^{f_0}$ &  0.3725 &5.711&  $a_0^{h_+}$ &  0.4960 &$5.416^{*}$\\
$a_1^{f_0}$ & -0.9389 &5.711&  $a_1^{h_+}$ & -1.1275 &$5.416^{*}$\\
$a_0^{f_\perp}$ & 0.5182 &$5.416^{*}$&  $a_0^{h_\perp}$ &  0.3876&$5.416^{*}$ \\
$a_1^{f_\perp}$ & -1.3495 &$5.416^{*}$&  $a_1^{h_\perp}$ & -0.9623&$5.416^{*}$ \\
$a_0^{g_\perp,g_+}$ &  0.3563 &5.750&  $a_0^{\widetilde{h}_{\perp},\widetilde{h}_+}$ &  0.3403&5.750 \\
$a_1^{g_+}$ & -1.0612 &5.750&  $a_1^{\widetilde{h}_+}$ & -0.7697 &5.750\\
$a_0^{g_0}$ &  0.4028 &$5.367^{*}$&  $a_1^{\widetilde{h}_\perp}$ & -0.8008&5.750 \\
\hline\hline
\end{tabular}
\caption{The central values for the form factors parameters $a_0^f$, and $a_1^f$ involved in the fit of equation (27) using the lattice QCD approach [35]. Values of the pole masses, $m_{\rm pole}^*$ are from the particle data group [47], while $m_{\rm pole}$ masses were taken from the lattice QCD calculations [48].}
\end{table}
\begin{table}[ht]
\centering
\begin{tabular}{l}
\hline\hline
$m_{\Lambda_{b}}=5.62$ GeV, $m_{b}=4.8$ GeV, $m_{\mu}=0.105$ GeV,\\
$m_{\tau}=1.77$ GeV, $|V_{tb}V_{ts}^{\ast}|=45\times10^{-3}$, $m_{c}=1.3$ GeV\\ $\alpha^{-1}=137$, $G_{F}=1.17\times 10^{-5}$ GeV$^{-2}$,\\
$\tau_{\Lambda_{b}}=1.383\times 10^{-12}$ sec, $m_{\Lambda}=1.115$ GeV,\\
\hline\hline
\end{tabular}
\caption{Values of input parameters used in our numerical analysis [47]. }
\end{table}

\clearpage
\begin{table}[ht]
\centering
\scalebox{0.9}{
\begin{tabular}{ccccc}
\hline
Branching Ratio$ \times 10^{-6}$  & $%
\begin{array}{c}
\Lambda _{b}\rightarrow \Lambda \mu ^{+}\mu ^{-} \\
\text{without LD}%
\end{array}%
$ & $%
\begin{array}{c}
\Lambda _{b}\rightarrow \Lambda \mu ^{+}\mu ^{-} \\
\text{with LD}%
\end{array}%
$ & $%
\begin{array}{c}
\Lambda _{b}\rightarrow \Lambda \tau ^{+}\tau ^{-} \\
\text{without LD}%
\end{array}%
$ & $%
\begin{array}{c}
\Lambda _{b}\rightarrow \Lambda \tau ^{+}\tau ^{-} \\
\text{with LD}%
\end{array}%
$ \\ \hline
SM & $1.9$ & $12.38$ & $0.48$ & $4.49$ \\ \hline
$m_{H}=160 GeV$ $\mid\lambda_{tt}\mid = 0.05$  & $2.35$     & $ 12.78$    & $ 0.57$ & $4.55$ \\ \hline
$m_{H}=160 GeV$ $\mid\lambda_{tt}\mid = 0.10$  & $3.47$    & $14.33$      & $0.71$ & $4.67$ \\ \hline
$m_{H}=160 GeV$ $\mid\lambda_{tt}\mid = 0.15$  & $5.26$    & $16.09$      & $0.96$ & $4.86$ \\ \hline
$m_{H}=200 GeV$ $\mid\lambda_{tt}\mid = 0.05$ & $2.34$      & $12.68$      & $0.55$ & $4.54$ \\ \hline
$m_{H}=200 GeV$ $\mid\lambda_{tt}\mid = 0.10$  &  $3.06$   & $13.95$      & $0.66$ & $4.63$ \\ \hline
$m_{H}=200 GeV$ $\mid\lambda_{tt}\mid = 0.15$  & $4.39$    & $15.21$      & $0.82$ & $4.77$ \\ \hline
$m_{H}=300 GeV$ $\mid\lambda_{tt}\mid = 0.05$  & $2.09$     & $12.54$     & $0.52$ & $4.52$ \\ \hline
$m_{H}=300 GeV$ $\mid\lambda_{tt}\mid = 0.10$ &  $2.51 $    & $12.93$     & $0.59$ & $4.57$ \\ \hline
$m_{H}=300 GeV$ $\mid\lambda_{tt}\mid = 0.15$  & $3.17$    & $14.04$      & $0.68$ & $4.65$ \\ \hline
$m_{H}=700 GeV$ $\mid\lambda_{tt}\mid = 0.05$ &  $1.95$     & $12.41$     & $0.49$ & $4.50$ \\ \hline
$m_{H}=700 GeV$ $\mid\lambda_{tt}\mid = 0.10$  & $2.02$     & $12.48$      & $0.52$ & $4.51$ \\\hline
$m_{H}=700 GeV$ $\mid\lambda_{tt}\mid = 0.15$  & $2.14$    & $12.58$      & $0.54$ & $4.53$ \\ \hline

\end{tabular}}
\caption{The obtained central values of the branching ratios for $\Lambda _{b}\rightarrow \Lambda
\ell^{+}\ell^{-}$ in both SM and 2HDM. The experimental branching ratio for $\Lambda _{b}\rightarrow \Lambda
\mu^{+}\mu^{-}$ in units of $10^{-6}$ are measured by the CDF BR($\Lambda_{b}\to \Lambda\mu^{+}\mu^{-}$)=$1.73\pm 0.42(stat)\pm 0.55(syst)$ [22], and  LHCb  BR($\Lambda_{b}\to \Lambda\mu^{+}\mu^{-}$)=$0.96\pm 0.16(stat)\pm 0.13(syst)\pm 0.21(norm)$ [23] Collaborations, respectively.}
\end{table}

\clearpage
\begin{figure}[h!]
\begin{center}
\begin{tabular}{ccc}
\vspace{-2cm} \includegraphics[scale=0.4]{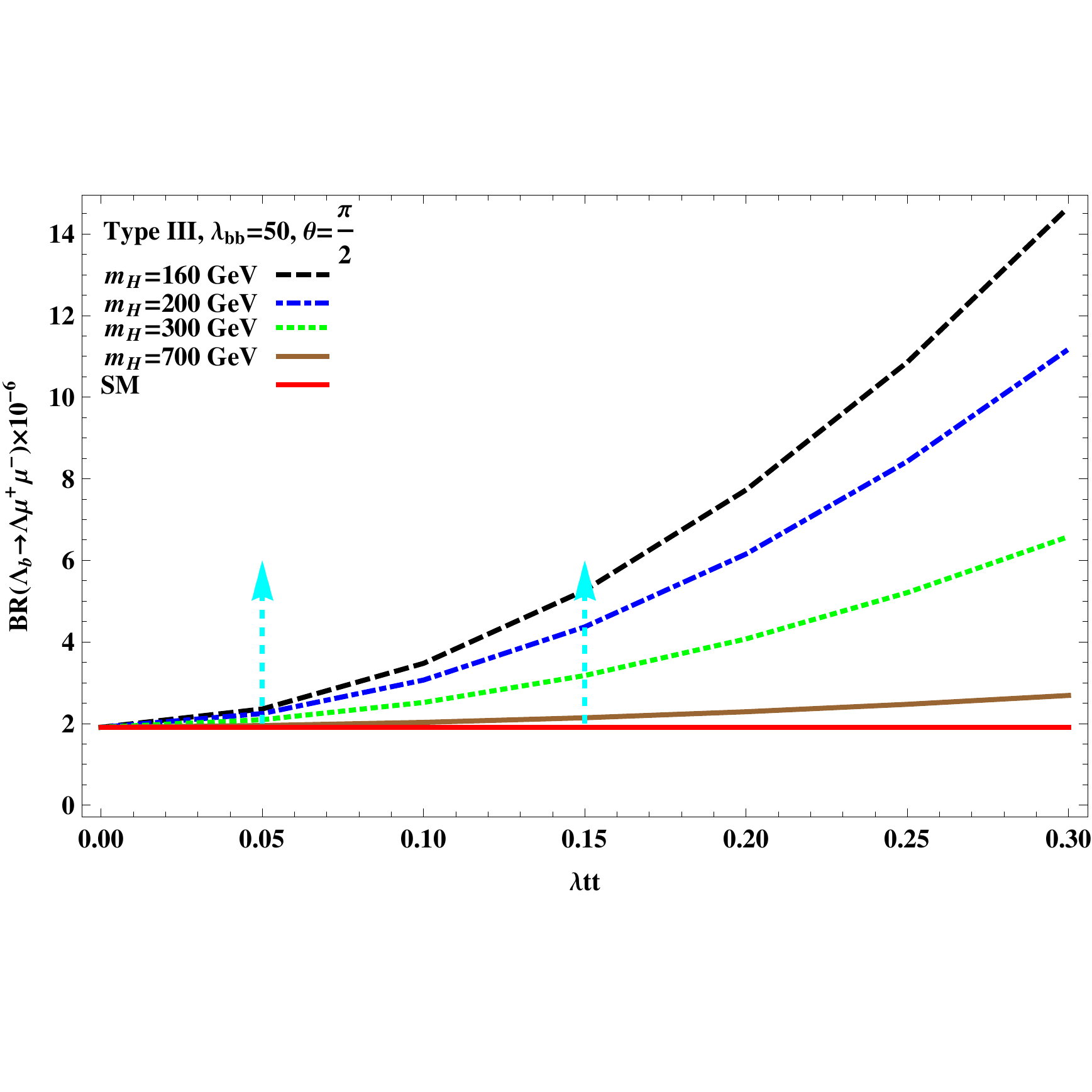}%
\includegraphics[scale=0.4]{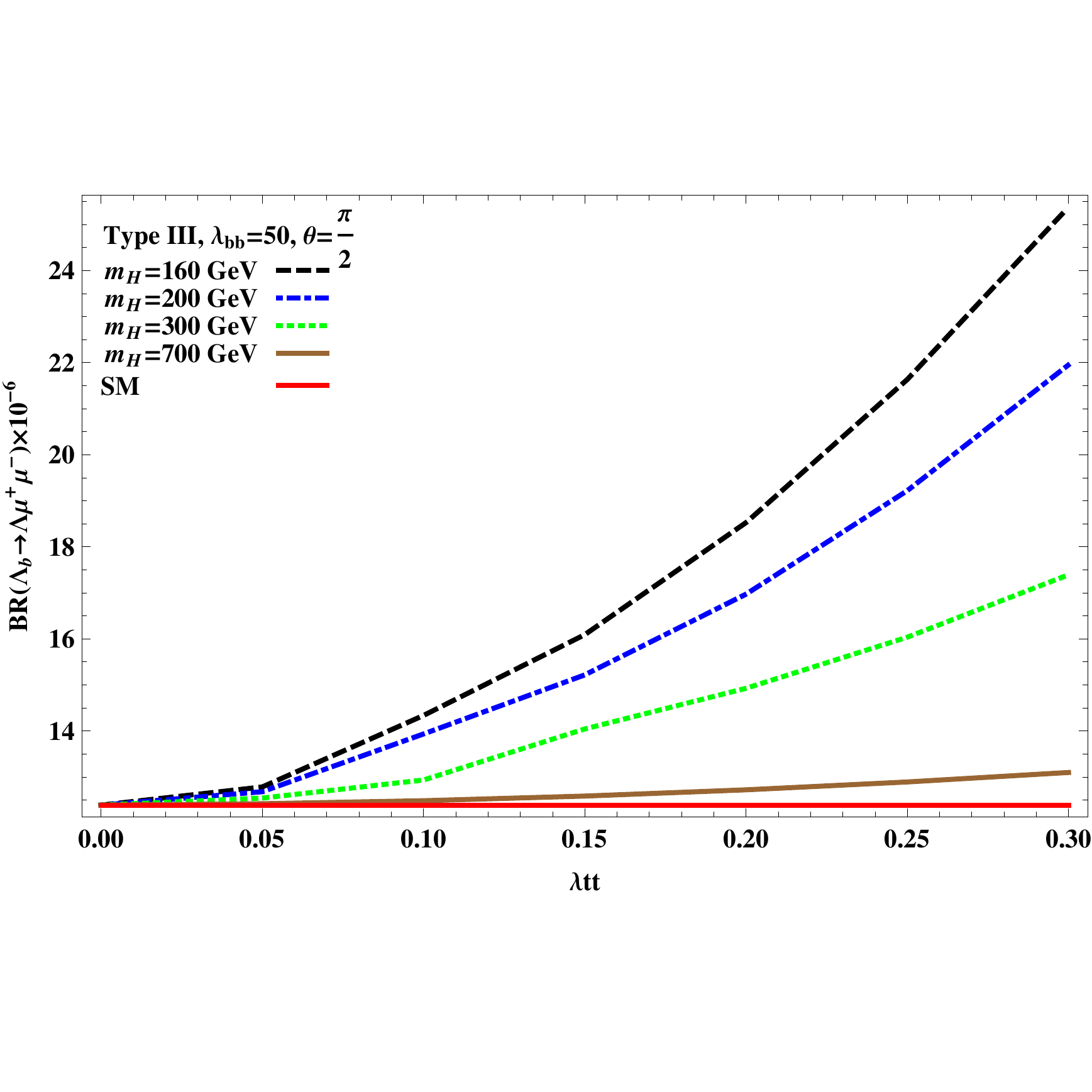} &  &  \\
\includegraphics[scale=0.4]{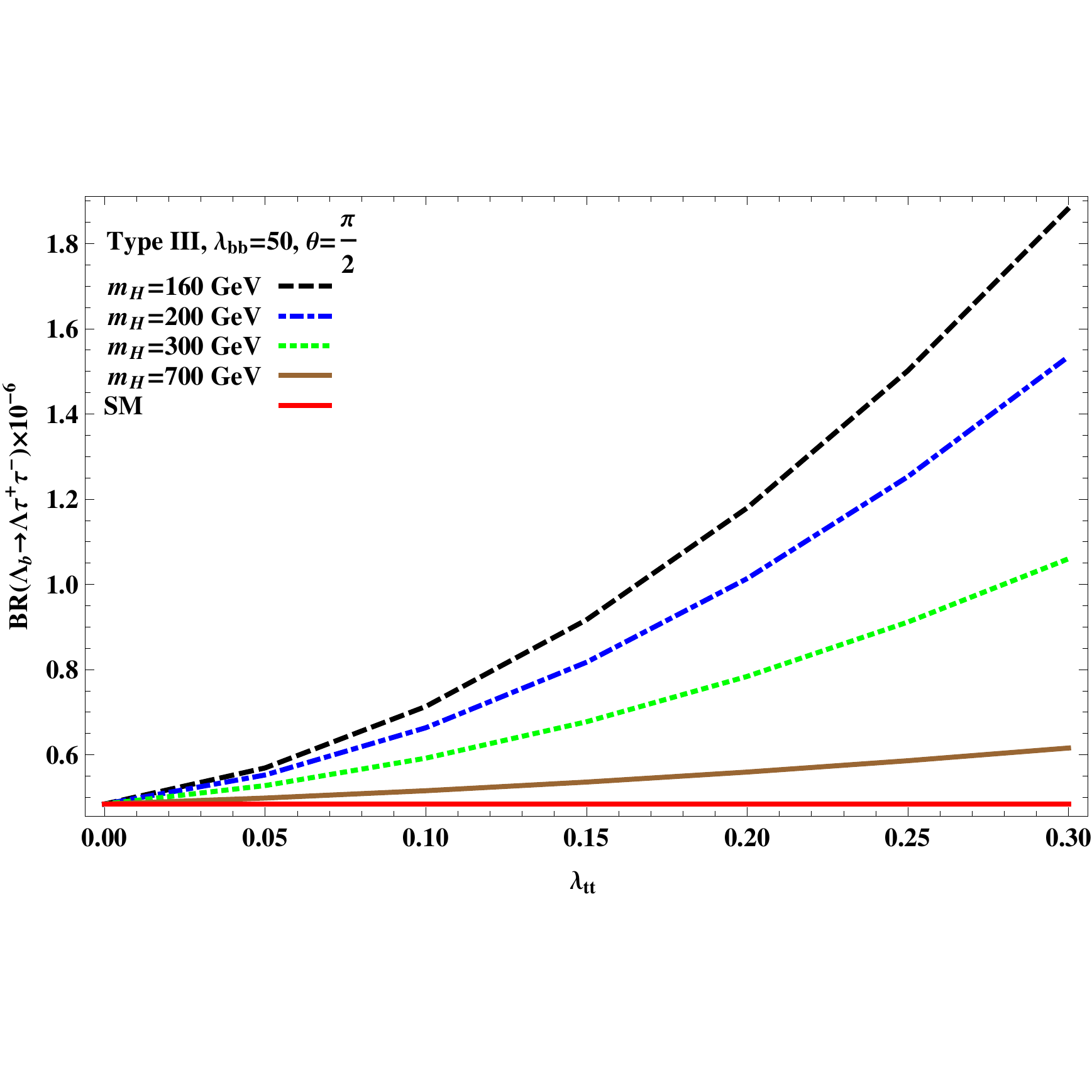}
\includegraphics[scale=0.4]{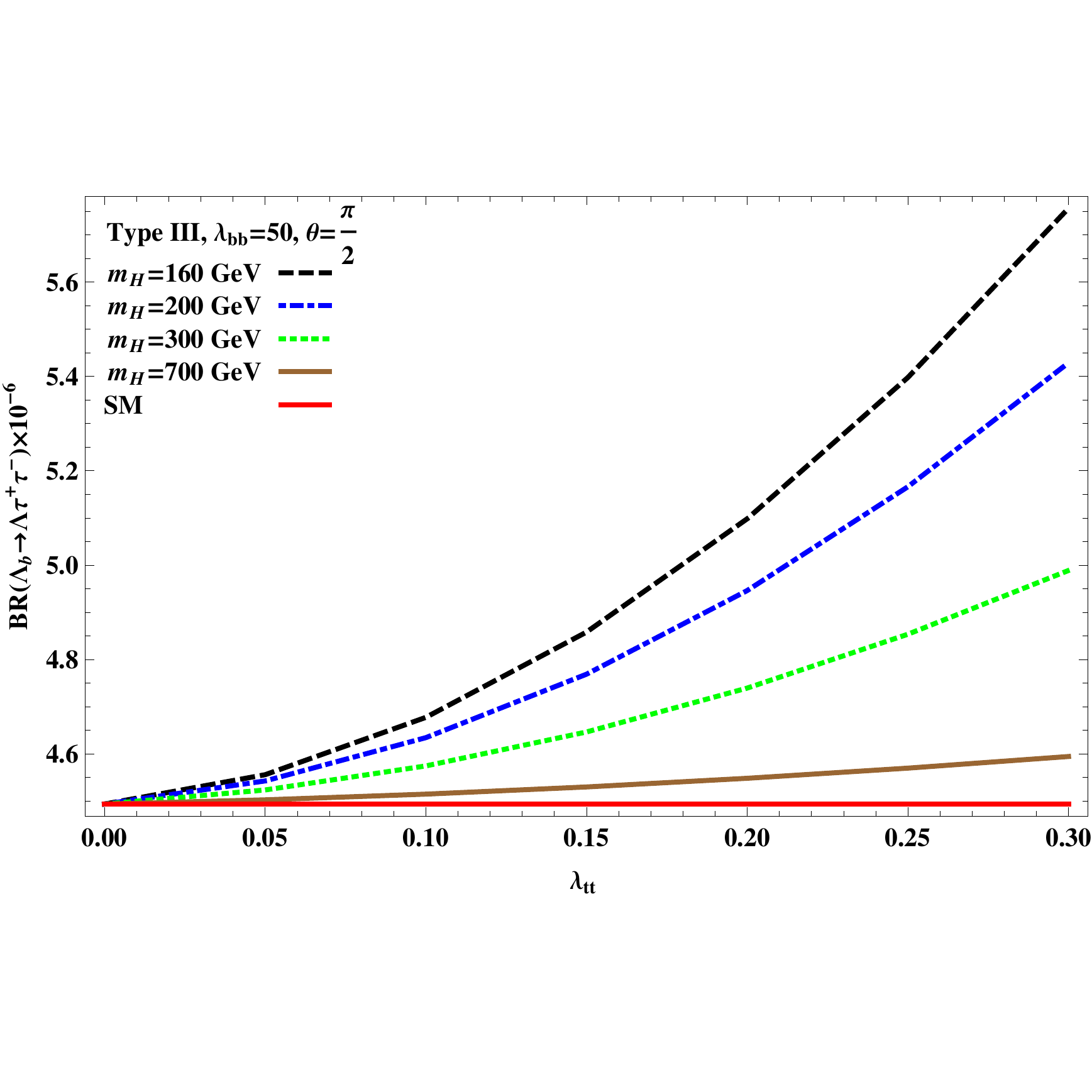}
\put (-290,172){(a)} \put (-90,172){(b)} \put (-295,25){(c)}
\put(-90,25){(d)} \vspace{-1cm} &  &
\end{tabular}%
\end{center}
\caption{The central values of branching ratios for the $\Lambda_b \to \Lambda \ell^+\ell^-$ ($\ell=%
\protect\mu, \protect\tau$) decays as functions of $\lambda_{tt}$ without
long-distance contributions (a, c) and with long-distance contributions (b,
d).}
\end{figure}
\begin{figure}[h!]
\begin{center}
\begin{tabular}{ccc}
\vspace{-2cm} \includegraphics[scale=0.4]{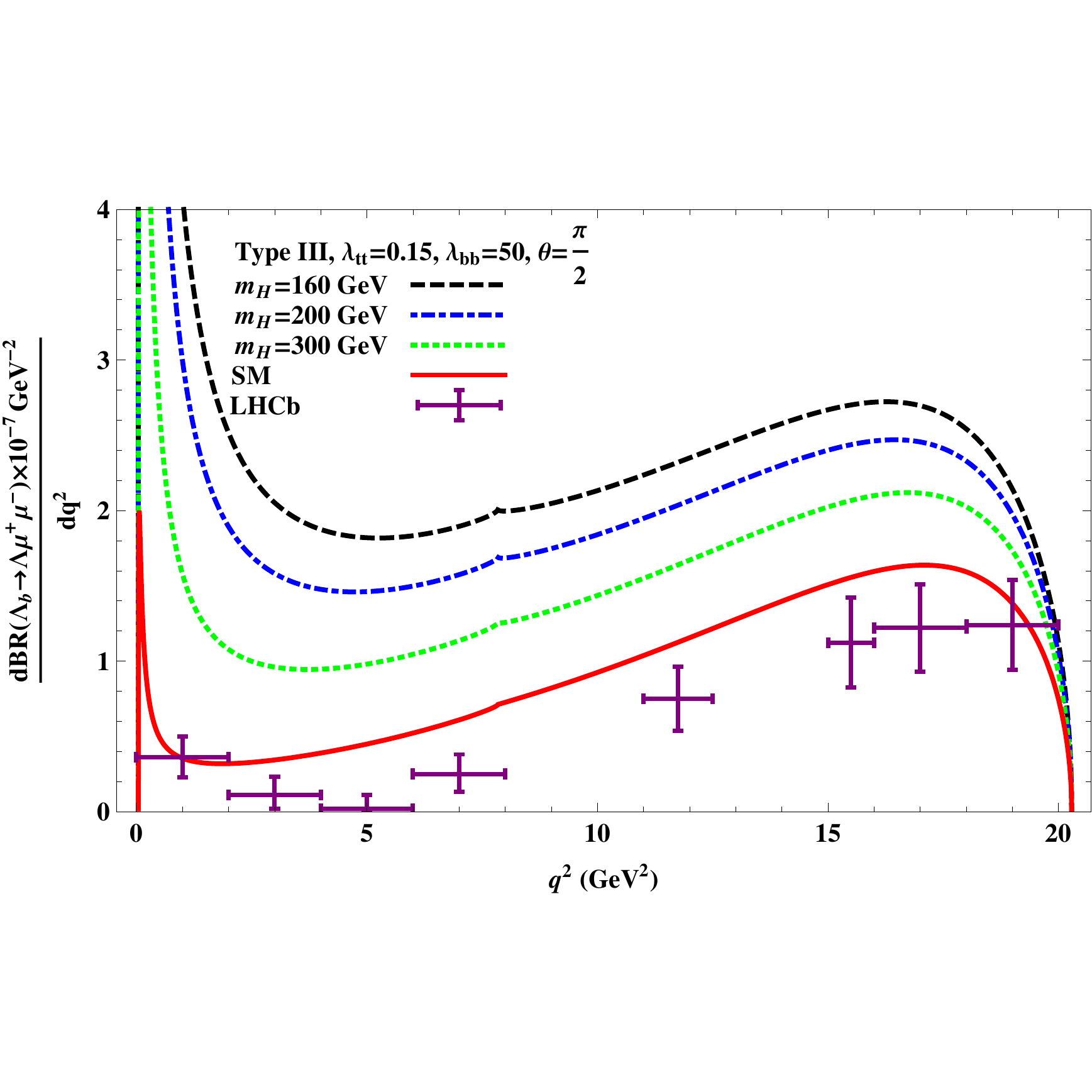} %
\includegraphics[scale=0.4]{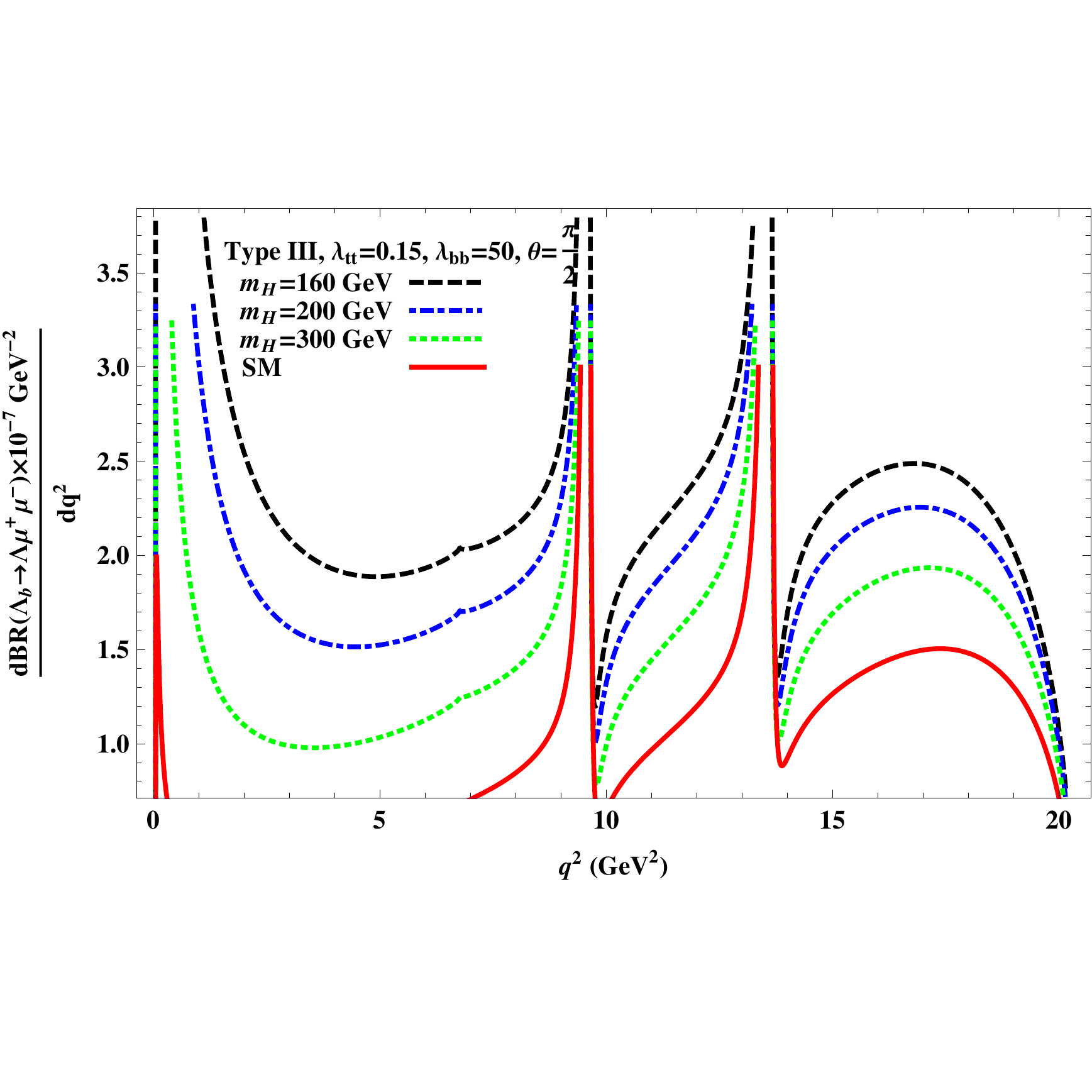} &  &  \\
\includegraphics[scale=0.4]{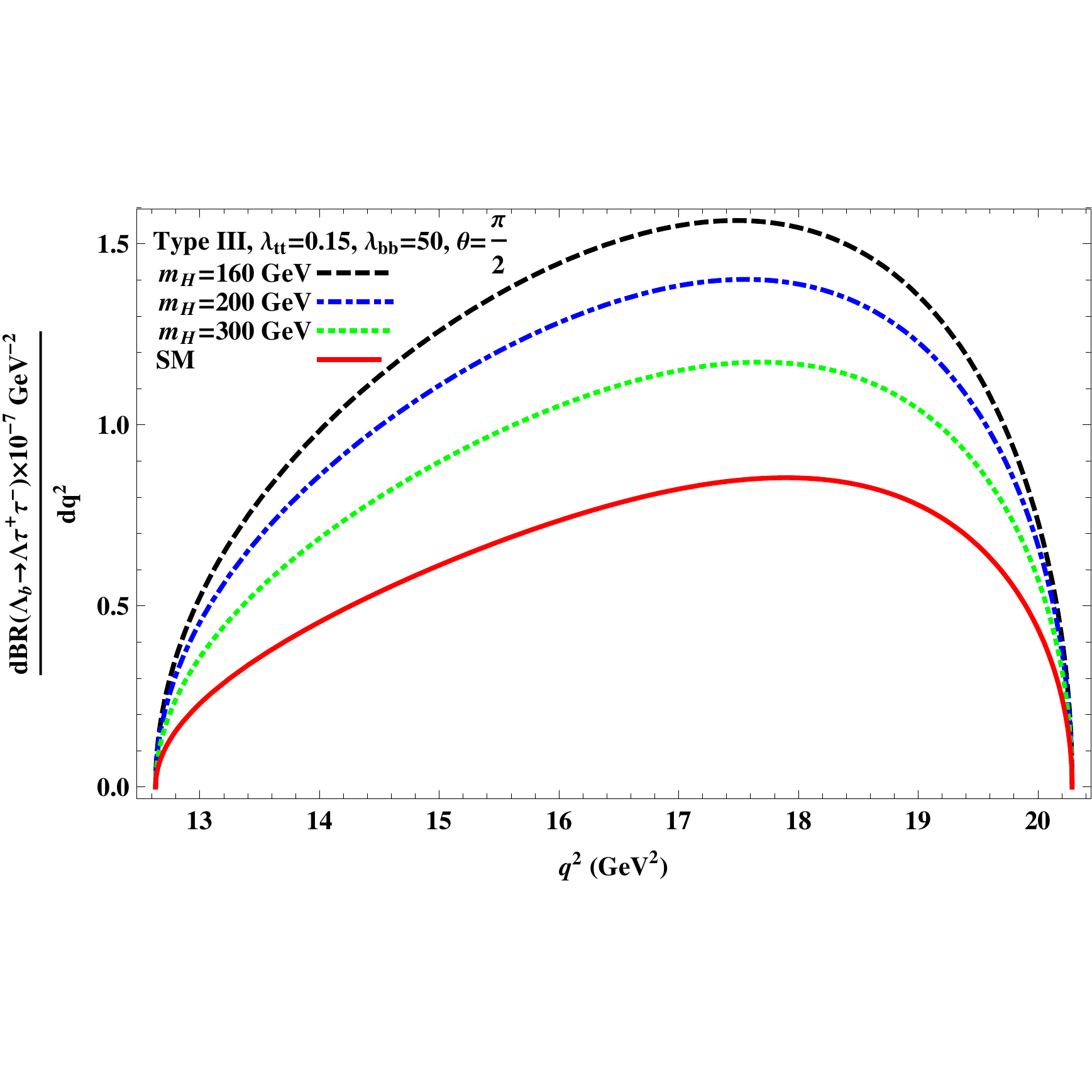} \includegraphics[scale=0.4]{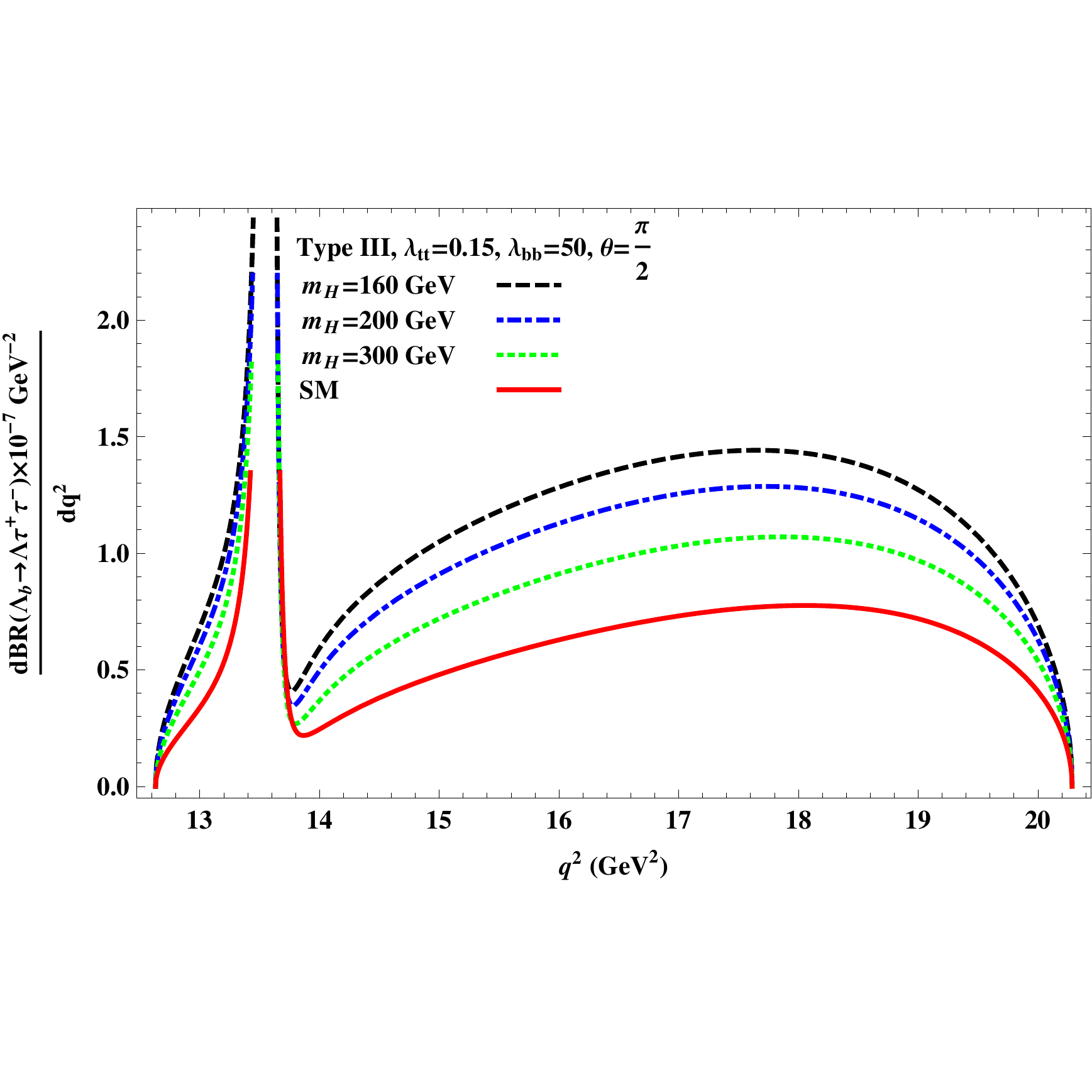}
\put (-289,175){(a)} \put (-85,175){(b)} \put (-292,25){(c)}
\put(-85,25){(d)} \vspace{-1cm} &  &
\end{tabular}%
\end{center}
\caption{The central values of the differential branching ratios for the $\Lambda_b \to \Lambda \ell^+\ell^-$ ($\ell=%
\protect\mu, \protect\tau$) decays as functions of $q^2$ without
long-distance contributions (a, c) and with long-distance contributions (b,
d). The solid (Red), dashed (Black), dashed-dot (Blue), and dotted (Green) lines represent, SM,
$m_{H }= 160~GeV$ , $ m_{H } = 200~GeV$ , and $m_{H } = 300~GeV$. The recent
experimental result LHCb [24] is also presented in (a).}
\end{figure}

\begin{figure}[h!]
\begin{center}
\begin{tabular}{ccc}
\vspace{-2cm} \includegraphics[scale=0.4]{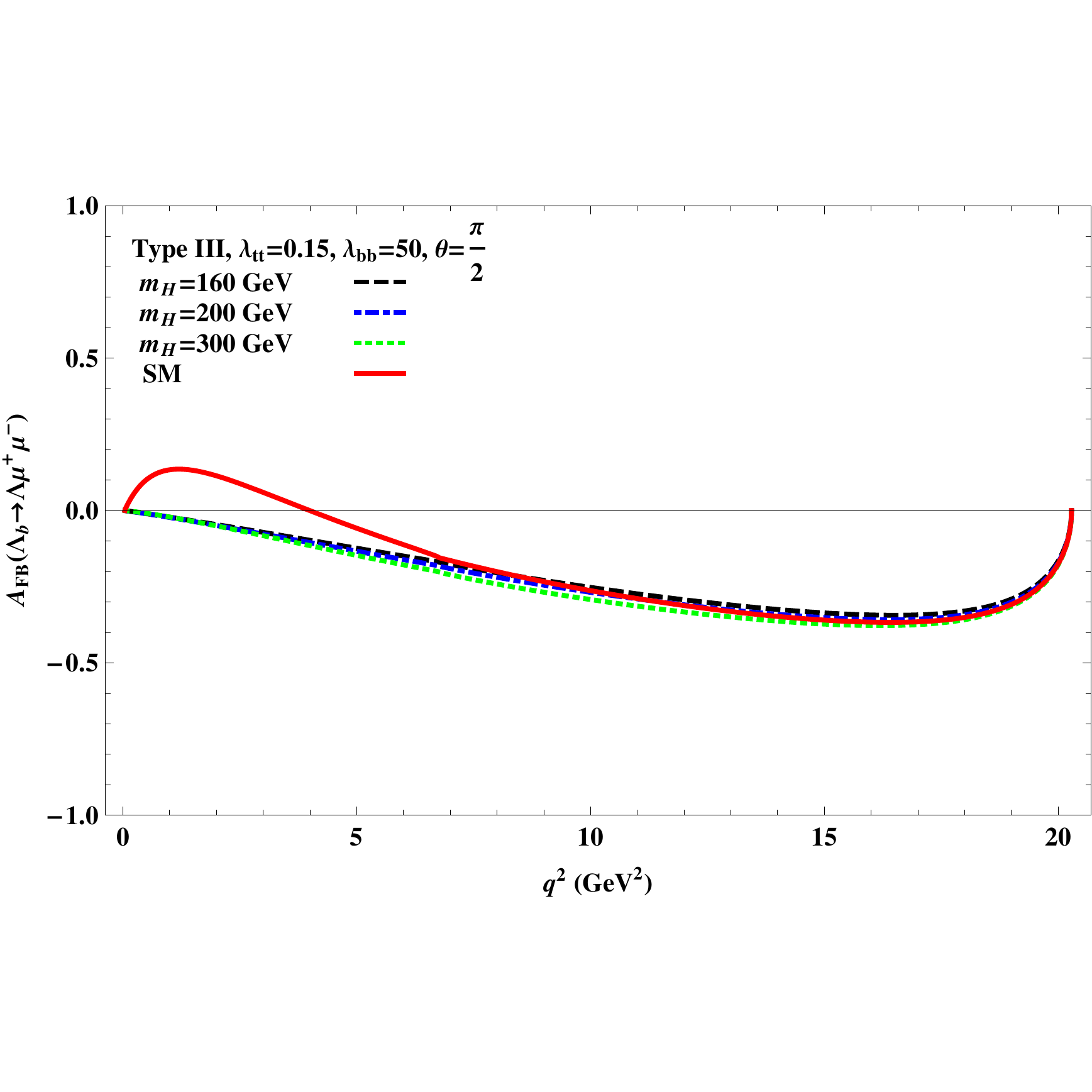} %
\includegraphics[scale=0.4]{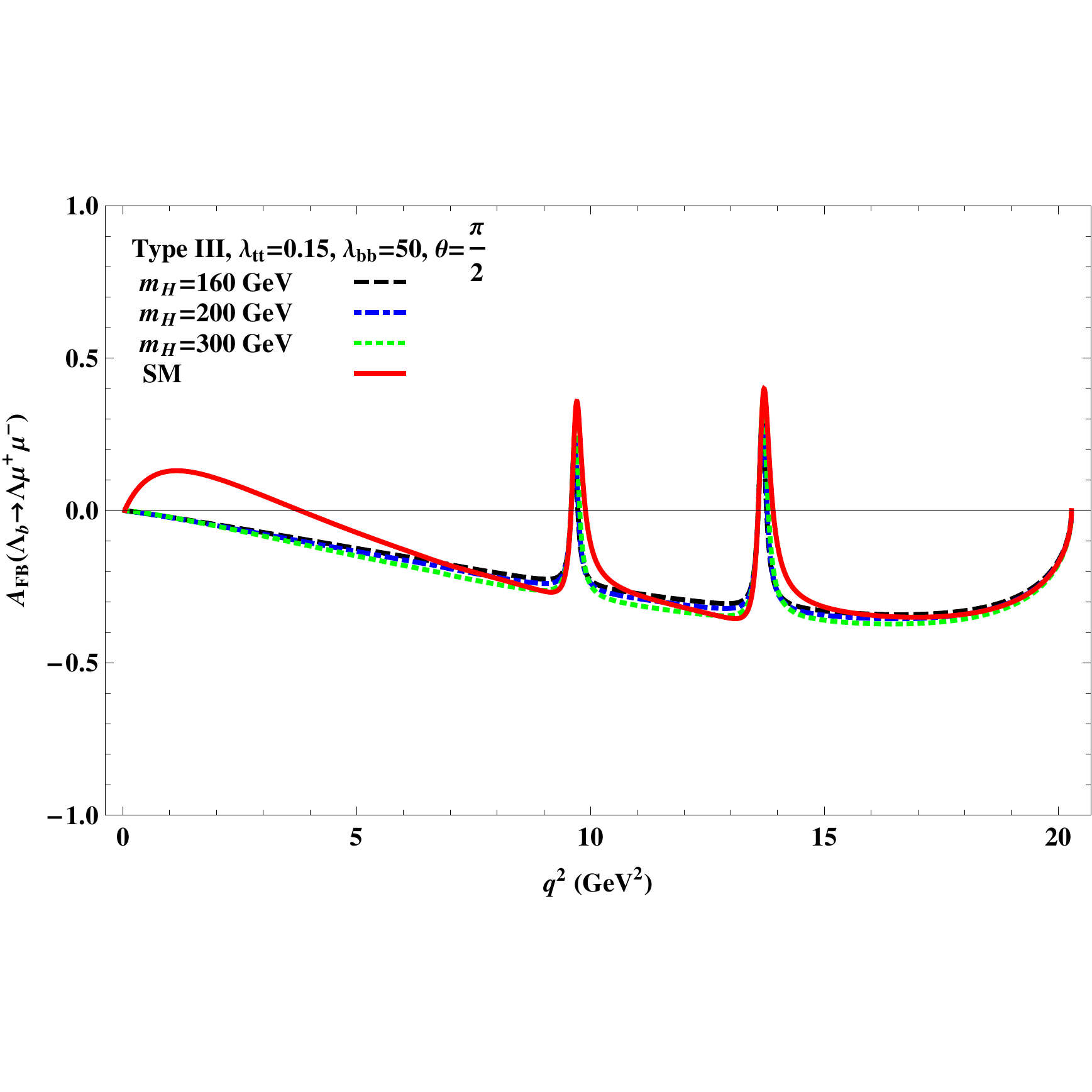} &  &  \\
\includegraphics[scale=0.4]{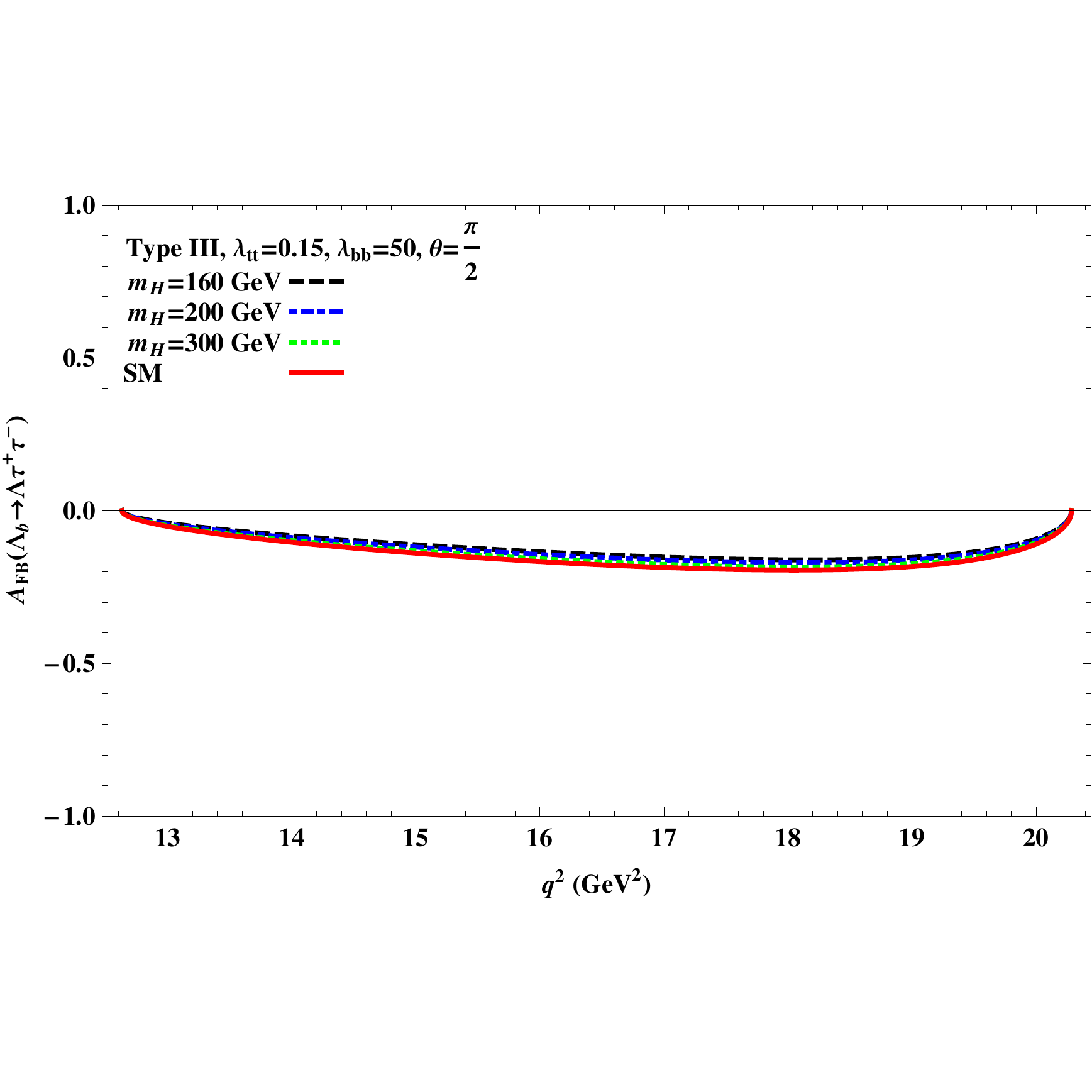} \includegraphics[scale=0.4]{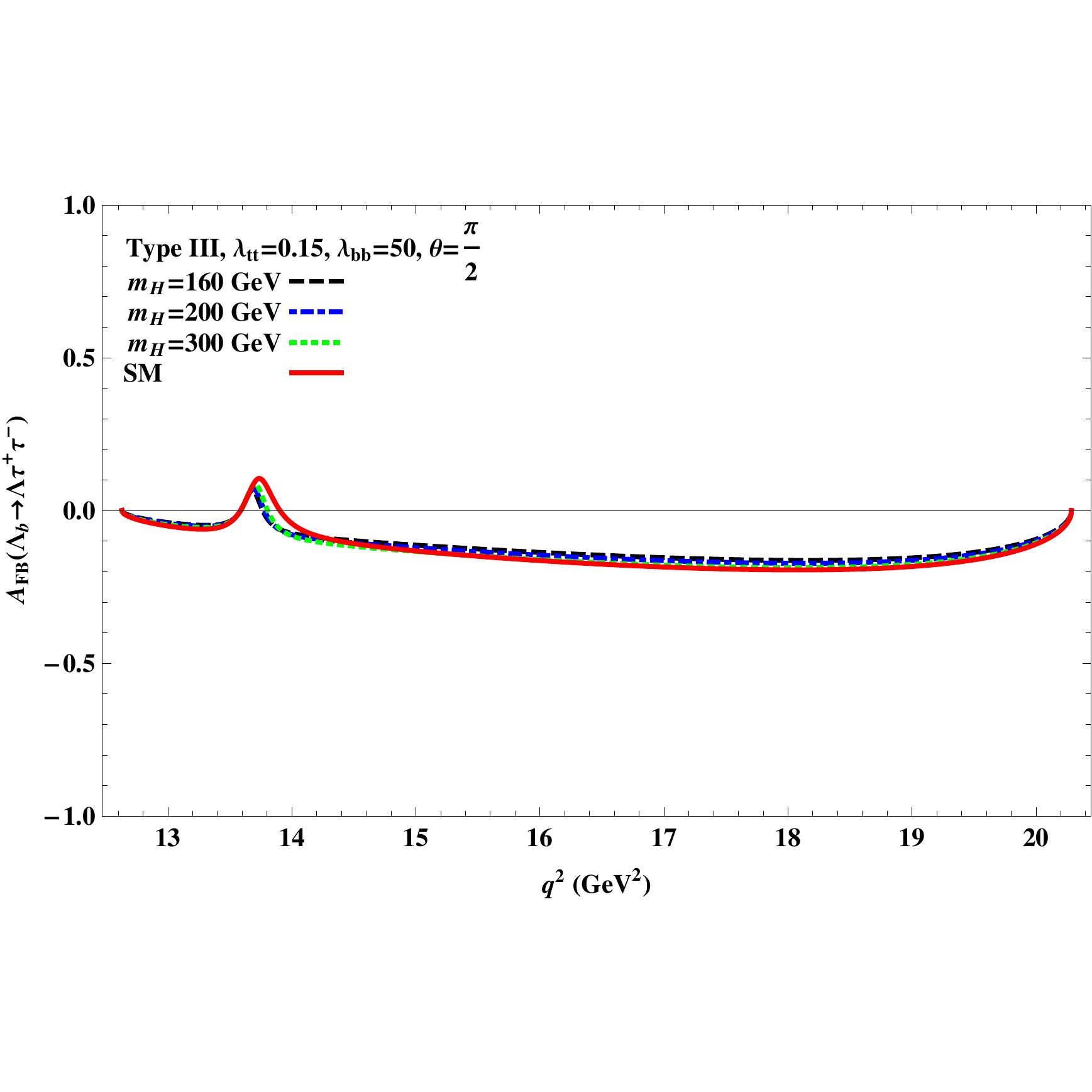}
\put (-298,172){(a)} \put (-94,172){(b)} \put (-300,22){(c)}
\put(-95,22){(d)} \vspace{-1cm} &  &
\end{tabular}%
\end{center}
\caption{The central values of the leptons forward-backward asymmetry for the $\Lambda_b \to \Lambda \ell^+\ell^-$ ($%
\ell=\protect\mu, \protect\tau$) decays as functions of $q^2$ without
long-distance contributions (a, c) and with long-distance contributions (b,
d). The line conventions are same as given in the legend of Figure 2. }
\end{figure}

\begin{figure}[h!]
\begin{center}
\begin{tabular}{ccc}
\vspace{-2cm} \includegraphics[scale=0.4]{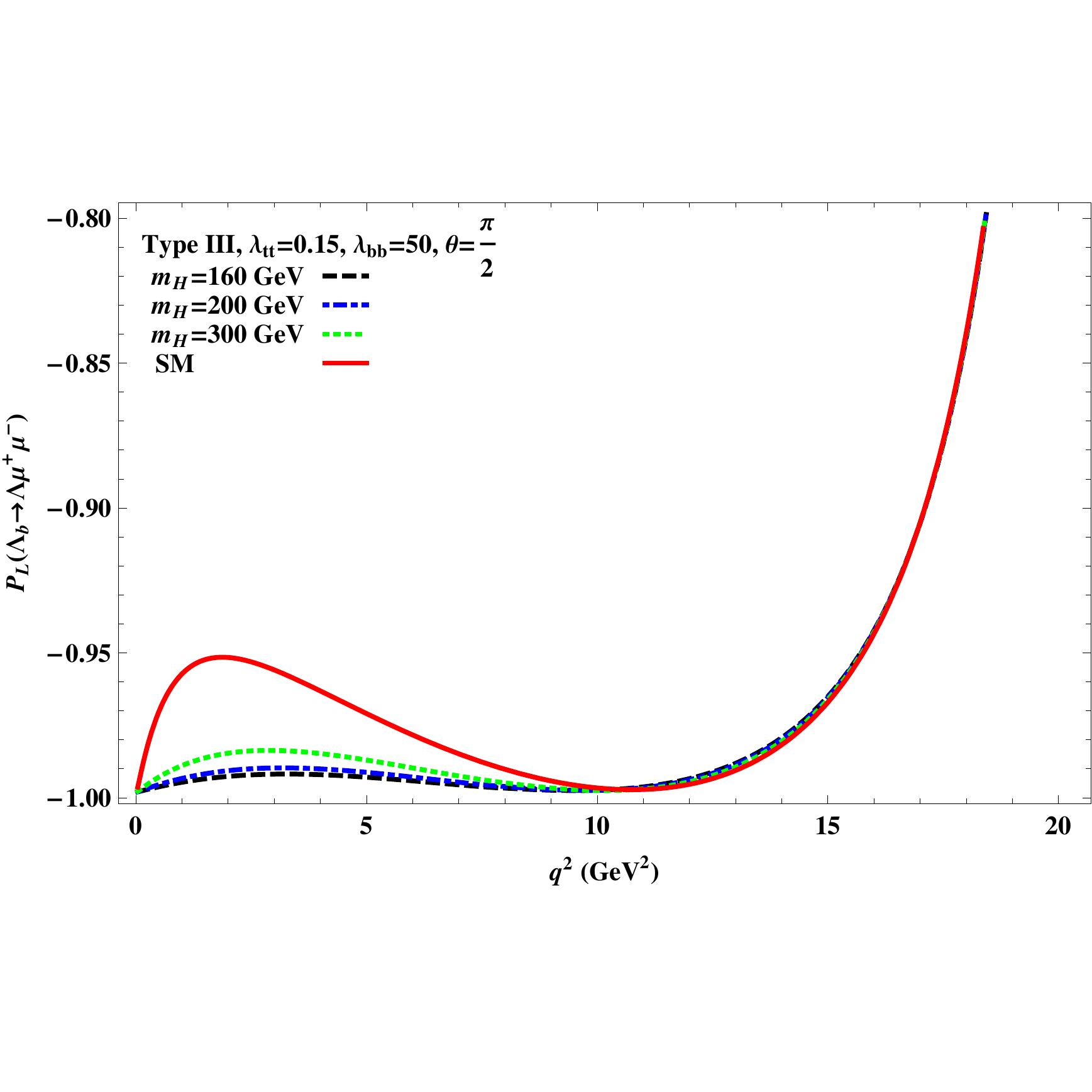} %
\includegraphics[scale=0.4]{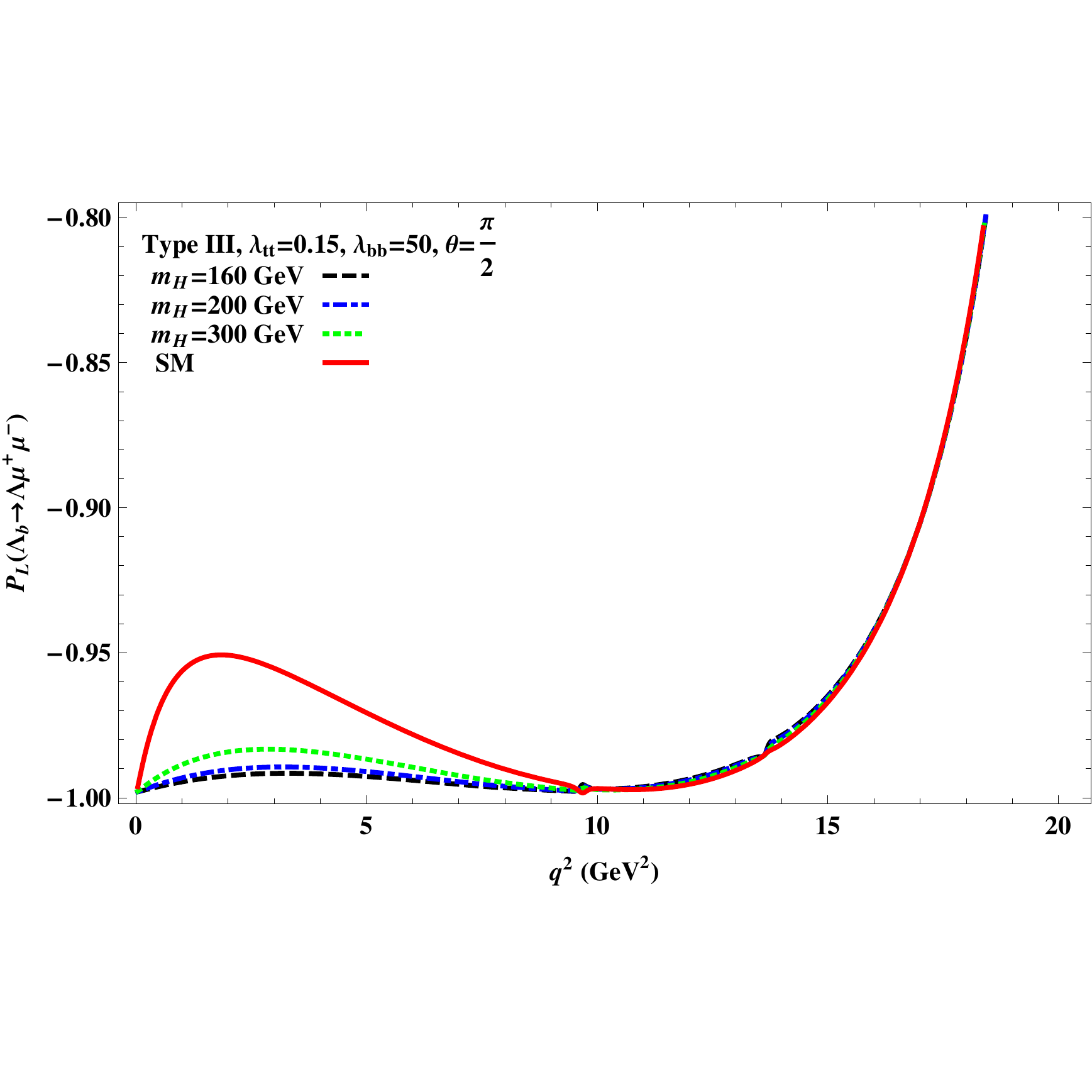} &  &  \\
\includegraphics[scale=0.4]{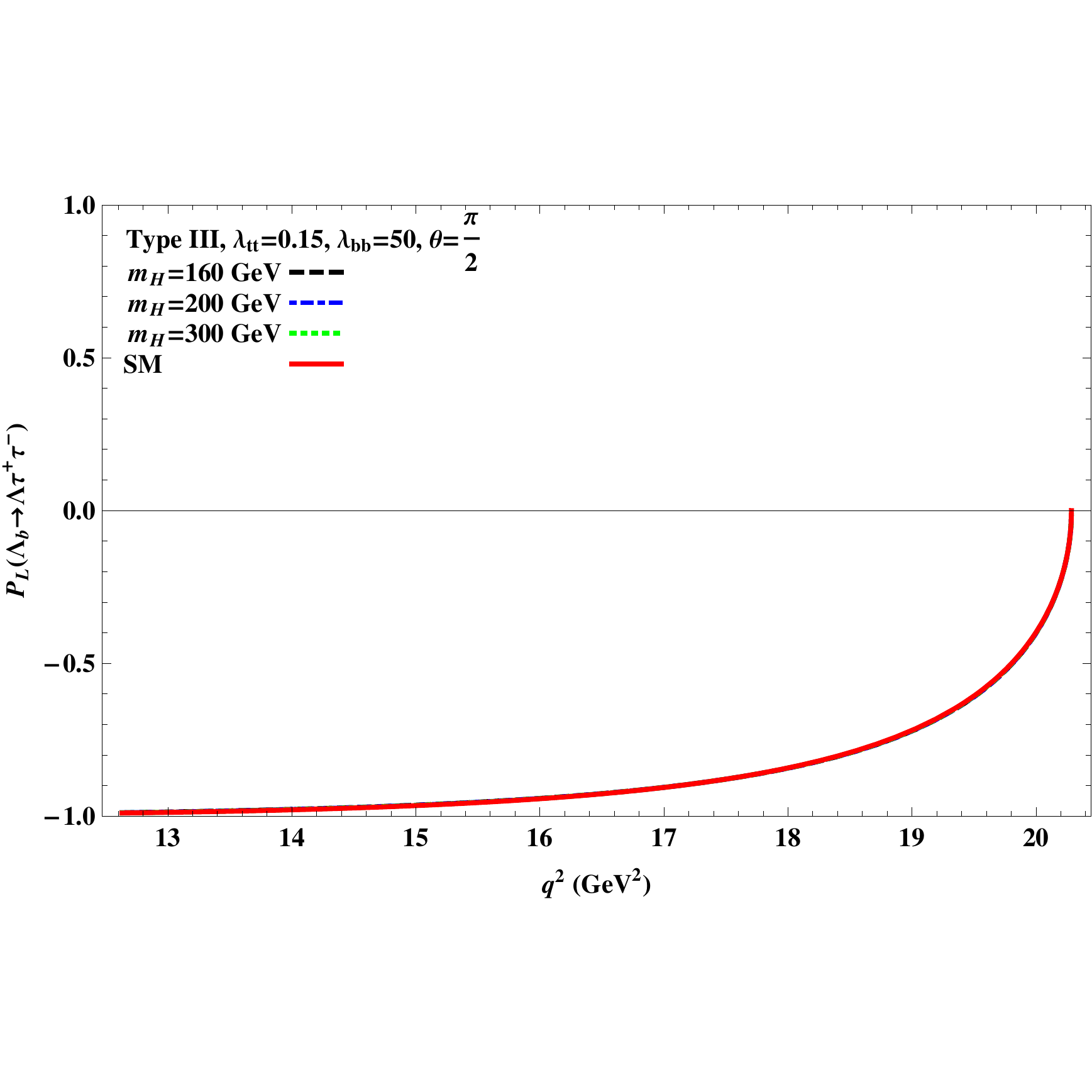} \includegraphics[scale=0.4]{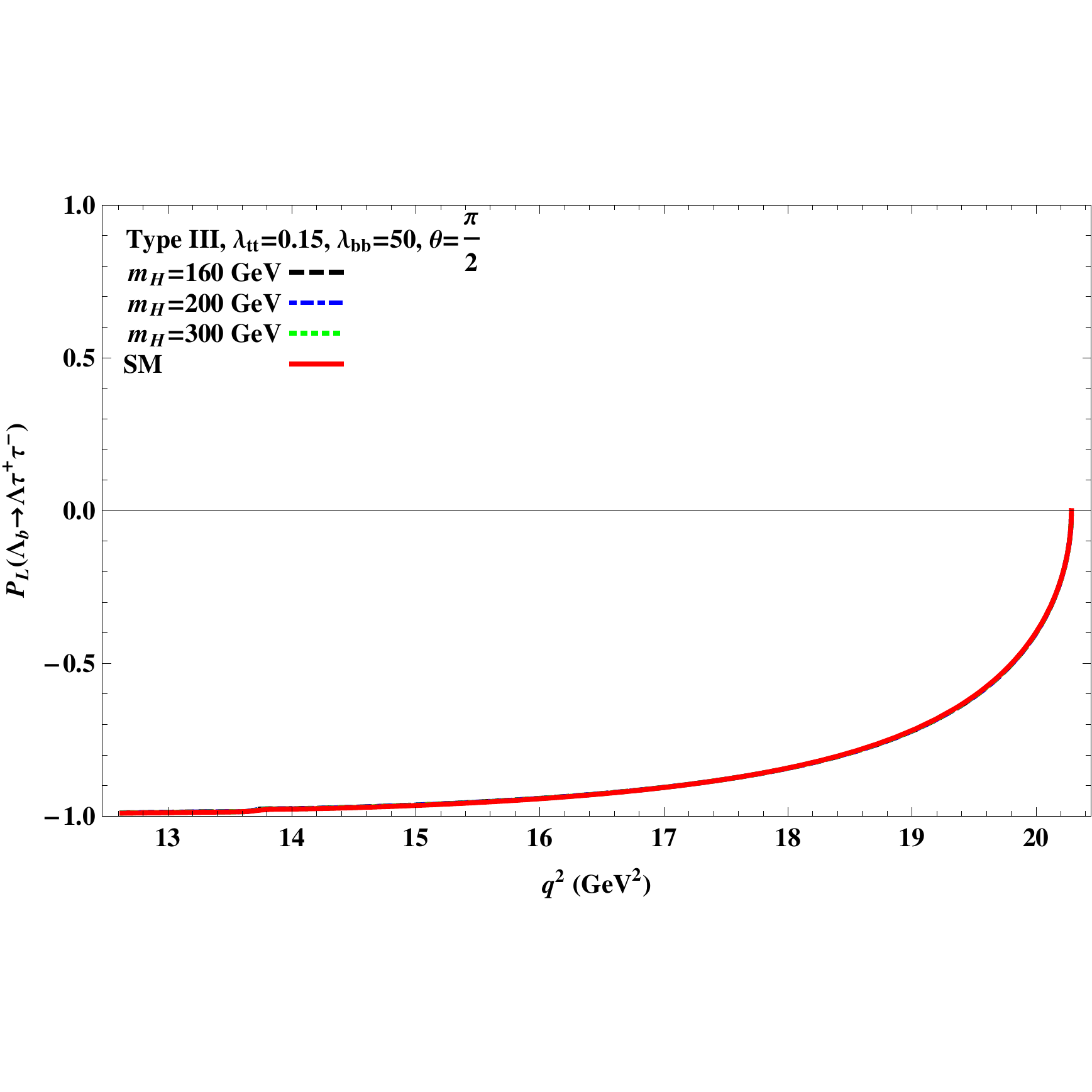}
\put (-298,172){(a)} \put (-95,172){(b)} \put (-300,25){(c)}
\put(-95,25){(d)} \vspace{-1cm} &  &
\end{tabular}%
\end{center}
\caption{The central values of the longitudinal $\Lambda$ polarization asymmetries for the
$\Lambda_b \to \Lambda \ell^+\ell^-$ ($\ell=\protect\mu, \protect\tau$)
decays as functions of $q^2$ without long-distance contributions (a,
c) and with long-distance contributions (b, d). The line conventions
are same as given in the legend of Figure 2.  }
\end{figure}

\begin{figure}[h!]
\begin{center}
\begin{tabular}{ccc}
\vspace{-2cm} \includegraphics[scale=0.4]{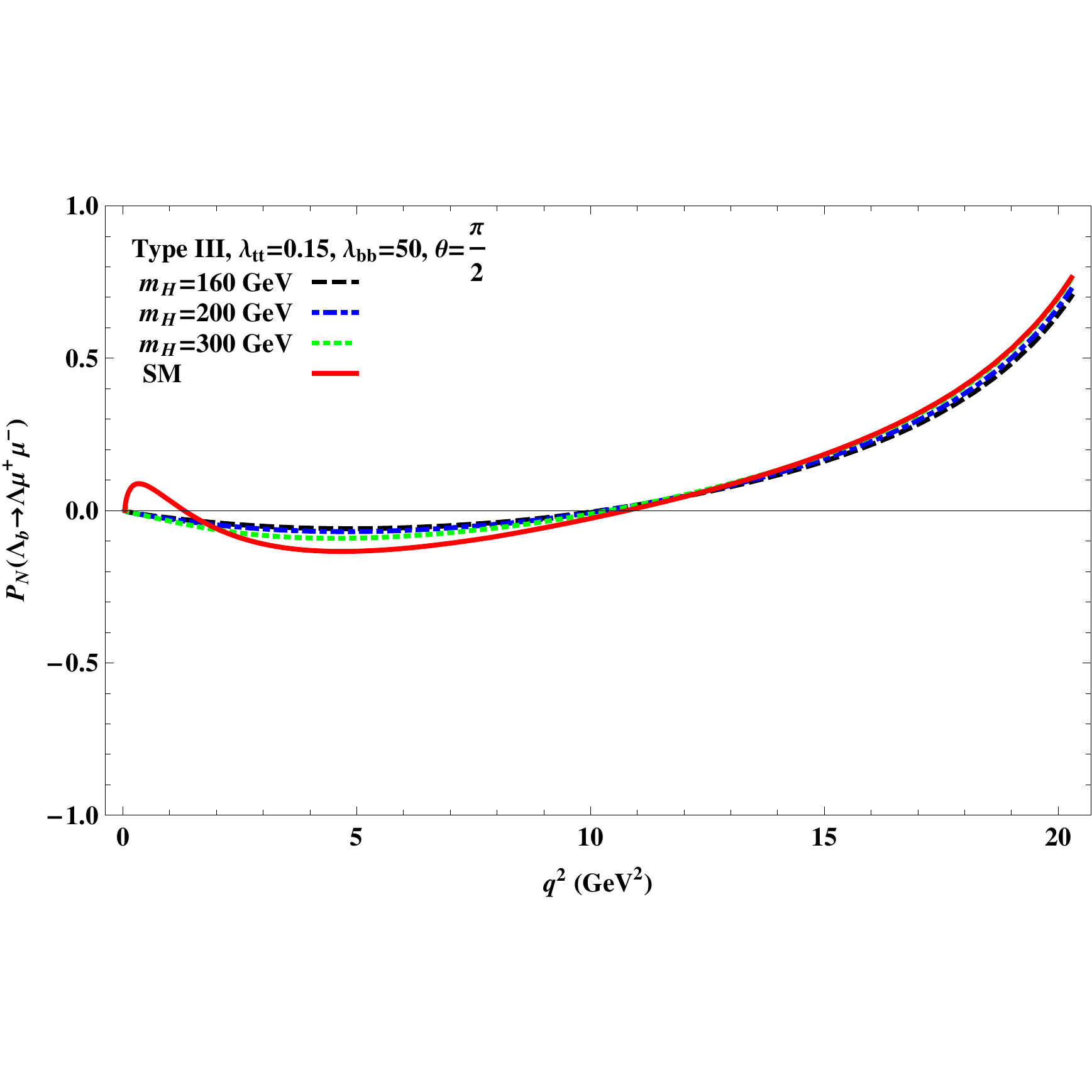} %
\includegraphics[scale=0.4]{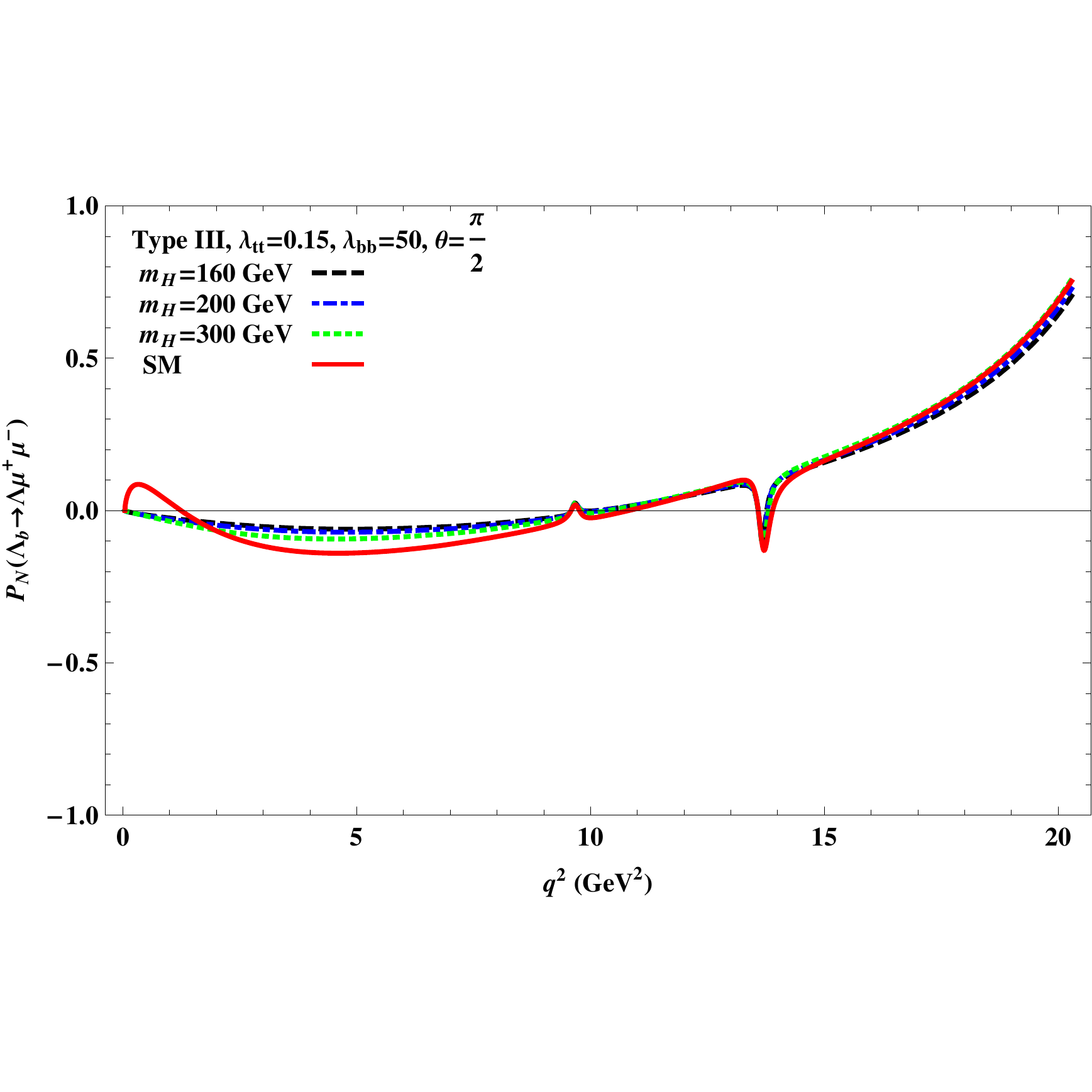} &  &  \\
\includegraphics[scale=0.4]{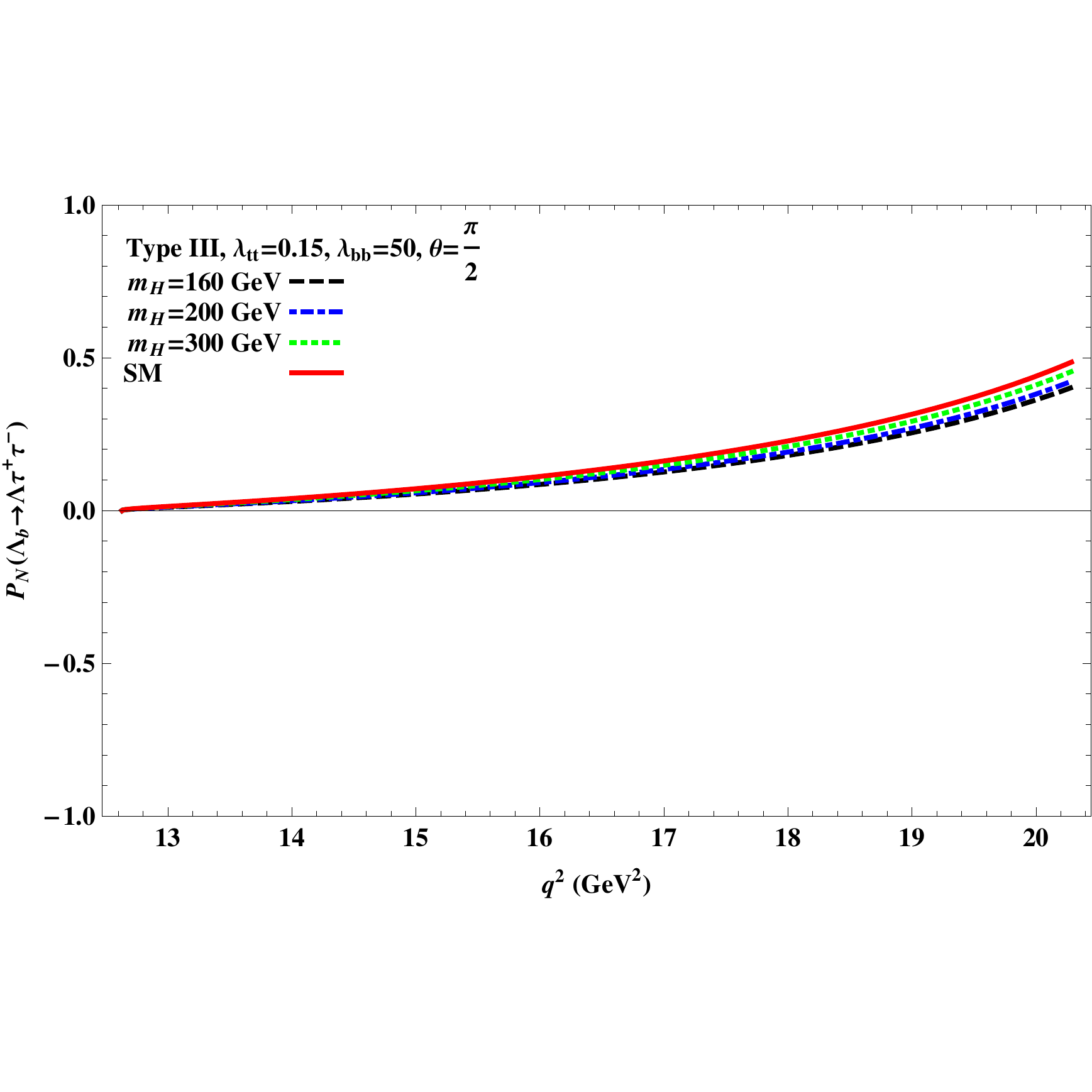} \includegraphics[scale=0.4]{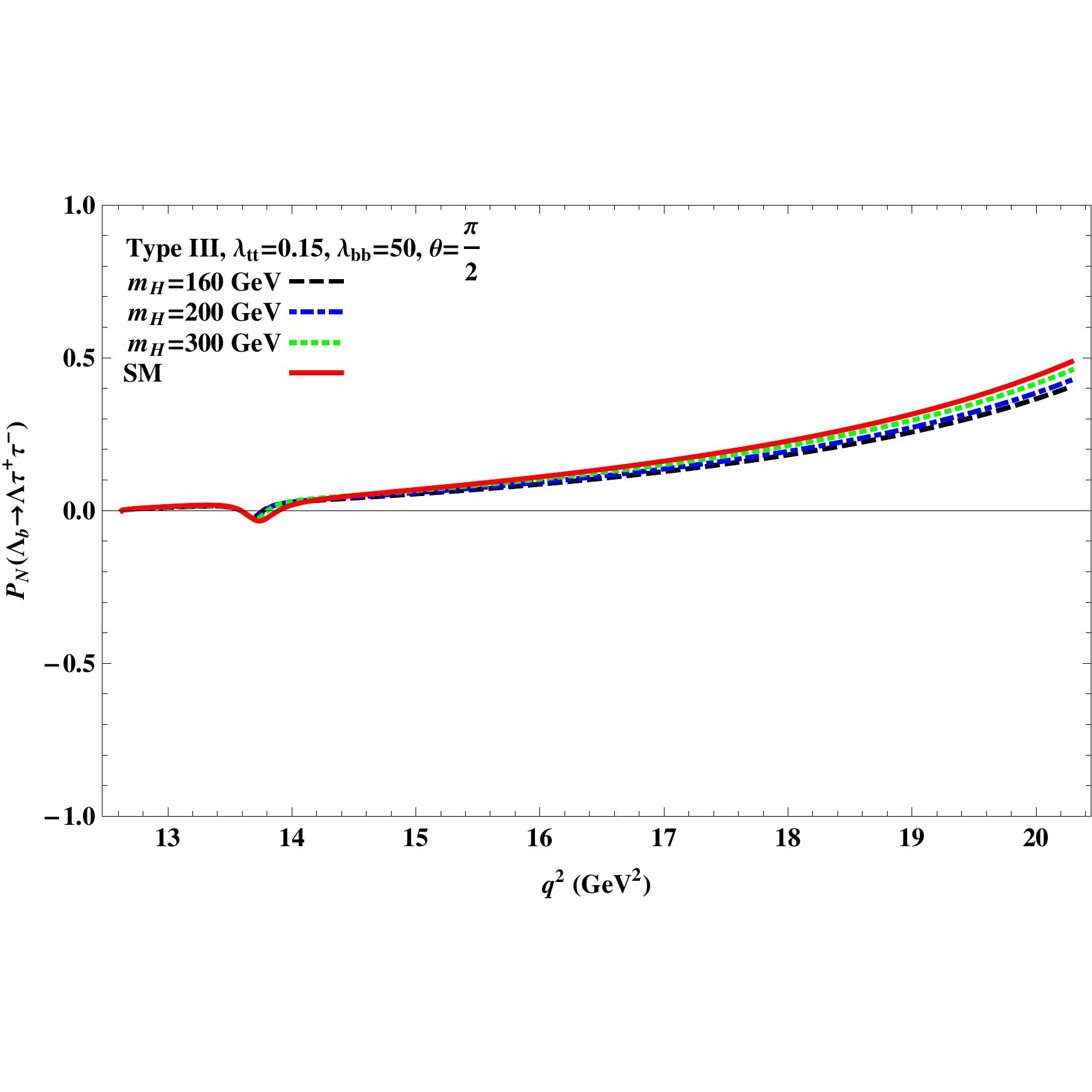}
\put (-298,172){(a)} \put (-95,172){(b)} \put (-300,22){(c)}
\put(-95,22){(d)} \vspace{-1cm} &  &
\end{tabular}%
\end{center}
\caption{The central values of the normal $\Lambda$ polarization asymmetries for the
$\Lambda_b \to \Lambda \ell^+\ell^-$ ($\ell=\protect\mu, \protect\tau$)
decays as functions of $q^2$ without long-distance contributions (a,
c) and with long-distance contributions (b, d). The line conventions
are same as given in the legend of Figure 2. }
\end{figure}


\clearpage
\begin{thebibliography}{99}
\bibitem{R1} ATLAS Collaboration, G. Aad et al., Phys. Lett. B 716, 1 (2012).
\bibitem{R2} CMS Collaboration, S. Chatrchyan et al., Phys. Lett. B 716, 30 (2012).
\bibitem{R3} A. Crivellin, C. Greub, and A. Kokulu, Phys. Rev. D 87, 094031 (2013).
\bibitem{R4} CMS Collaboration, S. Chatrchyan et al., JHEP 06, 081 (2013).
\bibitem{R5} T. Enomoto, and R. Watanabe, JHEP 05, 002 (2016).
\bibitem{R6} G. C. Branco, P. M. Ferreira, L. Lavoura, M. N. Rebelo, M. Sher, J. P. Silva, Phys. Report, 516, 1 (2012).
\bibitem{R7} W. Mader, J. H. Park, G. M. Pruna, D. St$\ddot{o}$ckinger, and A. Straessner, JHEP 09, 125 (2012); Erratum-JHEP 01, 006 (2014).
\bibitem{R8} LHCb Collaboration, R. Aaij et al.,  Phys. Rev. Lett. 111, 191801 (2013); Phys. Rev. Lett. 113, 151601 (2014); JHEP 02, 104 (2016).
\bibitem{R9} Belle Collaboration, A. Abdesselam et al., arXiv:1604.04042 [hep-ex]; Belle Collaboration, S. Wehle et al., arXiv:1612.05014 [hep-ex].
\bibitem{R10} B. Capdevila, A. Crivellin, S. Descotes-Genon, J. Matias, and J. Virto, arXiv:1704.05340 [hep-ph].
\bibitem{R11} W. Altmannshofer, P. Stangl, D. M. Straub, arXiv:1704.05435 [hep-ph].
\bibitem{R12} G. D'Amico, M. Nardecchia, P. Panci, F. Sannino, A. Strumia, R. Torre, and A. Urbano,	arXiv:1704.05438 [hep-ph].
\bibitem{R13} L. S. Geng, B. Grinstein, S. J$\ddot{a}$ger, J. M. Camalich, X. L. Ren, and R. X. Shi, arXiv:1704.05446 [hep-ph].
\bibitem{R14} G. Hiller, and I. Nisandzic, arXiv:1704.05444 [hep-ph].
\bibitem{R15} A. Celis, J. Fuentes-Martin, A. Vicente, and J. Virto, arXiv:1704.05672 [hep-ph].
\bibitem{R16} A. K. Alok, D. Kumar, J. Kumar, and R. Sharma, arXiv:1704.07347 [hep-ph].
\bibitem{R17} A. K. Alok, B. Bhattacharya, A. Datta, D. Kumar, J. Kumar, and D. London, arXiv:1704.07397 [hep-ph].
\bibitem{R18} W. Wang, and S. Zhao, arXiv:1704.08168 [hep-ph].
\bibitem{R19} T. Mannel and S. Recksiegel, J. Phys. G 24, 979 (1998).
\bibitem{R20} T. M.  Aliev, A. Ozpineci, and M. Savci, Nucl. Phys. B 709, 115 (2005).
\bibitem{R21} Y. M. Wang, M. J. Aslam, and C. D. Lu, Eur. Phys. J. C 59, 847 (2009).
\bibitem{R22} CDF Collaboration, T. Aaltonen et al., Phys. Rev. Lett. 107, 201802 (2011).
\bibitem{R23} LHCb Collaboration, R. Aaij et al., Phys. Lett. B 725, 25 (2013).
\bibitem{R24} LHCb Collaboration, R. Aaij et al., JHEP 06, 115 (2015), arXiv:1503.07138 [hep-ex].
\bibitem{R25} T. Barakat, J. Phys. G 24, 1903 (1998); Il Nuovo Cimento 112, 697 (1999).
\bibitem{R26} T. M. Aliev, and E. Iltan, Phys. Rev. D 58, 095014 (1998); J. Phys. G 25, 989 (1999).
\bibitem{R27} B. Grinstein, M. J. Savage, and M. B. Wise, Nucl. Phys. B 319, 217 (1989).
\bibitem{R28} Y. B. Dai, C. S. Huang, J. T. Li, and  W. J. Li, Phys. Rev. D 67, 096007 (2003).
\bibitem{R29} C. S. Kim, Y. W. Yoon, and X. Yuan, JHEP 12, 038 (2015).
\bibitem{R30} F. Falahati, and R. Khosravi, Phys. Rev. D 85, 075008 (2012).
\bibitem{R31} A. J. Buras, and M. M$\ddot{u}$nz, Phys. Rev. D 52, 186 (1995).
\bibitem{R32} D. Melikhov, N. Nikitin, and S. Simula, Phys. Lett. B 430, 332 (1998).
\bibitem{R33} T. M. Aliev, A. Ozpineci, and M. Savci, Nucl. Phys. B 649, 168 (2003).
\bibitem{R34} C. H. Chen, C. Q. Ceng, Phys. Rev. D 64, 074001 (2001).
\bibitem{R35} W. Detmold, S. Meinel, Phys. Rev. D 93, 074501 (2016).
\bibitem{R36} Y. M Wang, Y. Li, and C. D Lu, Eur. Phys. J. C 59, 861 (2009).
\bibitem{R37} Y. M. Wang and Y. L. Shen, JHEP 1602, 179 (2016).
\bibitem{R38} T. M. Aliev, K. Azizi, M. Savci, Phys. Rev. D 81, 056006 (2010).
\bibitem{R39} Y. Liu, L. L. Liu, X. H. Guo, arXiv:1503.06907 (2015).
\bibitem{R40} D. Atwood, L. Reina and A. Soni, Phys. Rev. D 55, 3156 (1997).
\bibitem{R41} D. Bowser-Chao, K. Cheung, and W. Y. Keung, Phys. Rev. D 59, 115006 (1999).
\bibitem{R42} C. S Huang, and S. H. Zhu, Phys. Rev. D 68, 114020 (2003).
\bibitem{R43} A. G. Akeroyd et al., Eur. Phys. J. C 77, 276 (2017).
\bibitem{R44} S. Meinel, and D. V. Dyki,  Phys. Rev. D 94, 013007 (2016).
\bibitem{R45} G. Kumar, and M. Mahajan, arXiv:1511.00935 (2015).
\bibitem{R46} P. B$\ddot{o}$er, T. Feldmann, and D. v. Dyk, JHEP 01, 155 (2015).
\bibitem{R47} Particle Data Group Collaboration, K. Olive et al., "Review of Particle Physics," Chin. Phys. C 38, 090001 (2014).
\bibitem{R48} C. B. Lang, D. Mohler, S. Prelovsek, and R. M. Woloshyn, Phys. Lett. B 750, 17 (2015), arXiv:1501.01646 [hep-lat].
\end{thebibliography}
\end{document}